\documentclass[12pt]{article}
\usepackage{epsfig,amsfonts,amssymb}
\usepackage{hyperref}
\usepackage{cite}
\input epsf.sty
\usepackage{cite}

\def\ZZZ{{\hbox{ Z\kern-1.6mm Z}}}
\def\RRR{{\hbox{ R\kern-2.4mm R}}}
\def\CCC{{\hbox{ C\kern-2.0mm C}}}
\def\zzz{{\hbox{z\kern-1mm z}}}

\newcommand{\nn}{\nonumber \\}

\newcommand{\qeq}{{\hbox{=\kern-2.3mm ? \kern.5mm }}}
\renewcommand{\qeq}{=}

\newcommand{\eps}{\epsilon}

\newcommand{\vp}{\varphi}

\newcommand{\LL}{{\cal L}}

\newcommand{\wt}{\widetilde}

\newcommand{\NN}{{\cal N}}

\newcommand{\SSS}{{\cal S}}

\newcommand{\be}{\begin{equation}}
\newcommand{\ee}{\end{equation}}
\newcommand{\ben}{\begin{eqnarray}\displaystyle}
\newcommand{\een}{\end{eqnarray}}

\newcommand{\refb}[1]{(\ref{#1})}
\newcommand{\p}{\partial}
\newcommand{\sectiono}[1]{\section{#1}\setcounter{equation}{0}}

\def\one{{\hbox{ 1\kern-.8mm l}}}
\def\zero{{\hbox{ 0\kern-1.5mm 0}}}

\newcommand{\bea}[1]{\begin{eqnarray}\label{#1} }
\newcommand{\eea}{\end{eqnarray}}

\newcommand{\aaa}{b}

\begin{document}

\baselineskip 24pt

\begin{center}
{\Large \bf  Logarithmic Corrections to Schwarzschild and Other
Non-extremal Black Hole Entropy in Different Dimensions}

\end{center}

\vskip .6cm
\medskip

\vspace*{4.0ex}

\baselineskip=18pt

\centerline{\large \rm Ashoke Sen}

\vspace*{4.0ex}

\centerline{\large \it Harish-Chandra Research Institute}
\centerline{\large \it  Chhatnag Road, Jhusi,
Allahabad 211019, India}

\vspace*{1.0ex}
\centerline{\small E-mail:  sen@mri.ernet.in}

\vspace*{5.0ex}

\centerline{\bf Abstract} \bigskip

Euclidean gravity method has been successful in computing logarithmic
corrections to extremal black hole entropy in terms of 
low energy data, and gives
results in perfect agreement with the microscopic results in string theory.
Motivated by this success we apply Euclidean gravity to compute logarithmic
corrections to the entropy of
various
non-extremal black holes in different dimensions,
taking special care of integration over the zero modes and keeping
track of the ensemble
in which the computation is done.
These results provide strong constraint on any ultraviolet completion of
the theory if the latter is able to give an independent
computation of the entropy of non-extremal black
holes from microscopic description. For Schwarzschild black holes
in four space-time dimensions
the macroscopic result seems to disagree with the existing result
in loop quantum gravity.

\vfill \eject

\baselineskip=18pt

\tableofcontents

\sectiono{Introduction and summary}  \label{sintro}

One of the tests of any theory of quantum gravity is a successful
comparison between the macroscopic and the microscopic prediction
of black hole entropy. This also provides us with a deep connection
between the infrared and the ultraviolet properties of gravity. The
leading contribution to the black hole entropy, given by the
Bekenstein-Hawking formula, can be computed 
from the low energy properties of gravity; yet it must
agree with the logarithm of the microstate degeneracy which is
sensitive to the ultraviolet completion of the theory.

Another property of the black hole entropy that can be computed
from the knowledge of the infrared physics is the logarithmic
correction to the black hole entropy.
By taking appropriate scaling limit of the mass, charge and other
quantum numbers carried by the black hole one can ensure that the
size of the black hole becomes large, but other dimensionless
ratios remain fixed. In this case the dominant contribution to the entropy
comes from the
Bekenstein-Hawking term, but it can receive subleading corrections
proportional to the logarithm of the horizon 
area\cite{9407001,9408068,
9412161,9510007,9604118,
9709064,0005017,
0104010,0112041,0406044,0409024,0805.2220,
0808.3688,0809.1508,0911.4379,1003.1083,1008.4314}.
On the macroscopic side
these corrections arise from one loop contribution to the
black hole partition function. Computation of 
the full one loop contribution would
certainly require knowledge of the ultraviolet completion of the theory,
but the logarithmic corrections arise only from loops of massless fields
and from the range of loop momentum
integration where the loop momenta remain 
much smaller than the Planck scale. Thus this can be evaluated
purely from the knowledge of the low energy data -- the spectrum
of massless fields and their coupling to the black hole background.
Requiring the
microscopic counting results to agree with this would give strong
constraint on any proposal for the ultraviolet completion of gravity.

\begin{table} {\small
\begin{center}\def\st{\vrule height 3ex width 0ex}
\begin{tabular}{|l|l|l|l|l|l|l|l|l|l|l|} \hline
The theory  & scaling of charges & logarithmic contribution & microscopic
\st\\[1ex] \hline \hline
$\NN=4$ supersymmetric CHL 
&  ${  Q_i}\sim \Lambda$,
\quad A $\sim\Lambda^2$  &   0 & $\surd$ \st\\[0 ex]
models in $D=4$ and type II on  &&& \st\\[0 ex]
$K3\times T^2$ with $  n_v$ matter multiplet &&& 
\st\\[1ex] \hline
Type II on $T^6$&   ${  Q_i}\sim \Lambda$, ${  A}\sim \Lambda^2$
& $-{  8\, \ln}\, \Lambda$ & $\surd$
\st\\[1ex] 
\hline
$\NN=2$ supersymmetric theories
 &   ${ 
Q_i}\sim\Lambda$, \quad A $\sim\Lambda^2$
&  ${  {1\over 6} (23 + n_H - n_V)\, \ln}\,  \Lambda$
& ?\st\\[0 ex] in $D=4$ with $  n_V$ vector and &&&
\st\\[0 ex]   $  n_H$  hyper multiplets & & &
\st\\[1ex] \hline
$\NN=6$ supersymmetric theories
&  ${  Q_i}\sim \Lambda$, \quad A $\sim\Lambda^2$
 & ${  -4\, \ln} \, \Lambda$ & ?  \st\\[0ex]
 in $D=4$ &&& \st\\[1ex] \hline
$\NN=5$ supersymmetric theories &  
${  Q_i}\sim \Lambda$, \quad A $\sim\Lambda^2$
 & ${  -2\, \ln} \, \Lambda$ & ? \st\\[0ex]
 in $D=4$ &&&
\st\\[1ex] \hline
$\NN=3$ supersymmetric theories in &  ${  Q_i}\sim \Lambda$, 
\quad A $\sim\Lambda^2$ & 
 ${  2\, \ln} \, \Lambda$ & ? \st\\[0 ex]
$D=4$ with $  n_v$ matter multiplets &&& \st\\[1ex] \hline
BMPV in type IIB on $T^5/\ZZZ_N$ &  ${  Q_1, Q_5, n}\sim \Lambda$, & 
 ${  -{1\over 4} (n_V+3) \, \ln} \, \Lambda$ & $\surd$
\st\\[0 ex] 
or $  K3\times S^1/\ZZZ_N$ with $  n_V$ vectors & ${  J=0}$, 
\quad A $\sim\Lambda^{3/2}$
& & \st\\[0ex]
preserving 16 or 32 supercharges &&& 
\st\\[1ex] \hline
BMPV in type IIB on $  T^5/\ZZZ_N$ &  ${  Q_1, Q_5, n}\sim \Lambda$, & 
 ${  -{1\over 4} (n_V-3) \, \ln} \, \Lambda$ & $\surd$
\st\\[0 ex] 
or $  K3\times S^1/\ZZZ_N$ with $  n_V$ vectors & 
${  J}\sim \Lambda^{3/2}$, \quad A $\sim\Lambda^{3/2}$
& & \st\\[0ex]
preserving 16 or 32 supercharges &&& 
\st\\[1ex] \hline
\hline
\end{tabular}
\end{center}
\caption{\small
Macroscopic predictions for the logarithmic
corrections to extremal black hole entropy in different 
string theories and
the status of their comparison with the microscopic results. 
The first column describes the theory and the black hole
under consideration. 
The second column describes the scaling of the various
charges, as well as  the area $A$ of the event horizon,
under which the logarithmic correction is computed.
For all four dimensional theories, 
$Q_i$ in the second column stands
for all the electric and magnetic charges of the black hole. For BMPV black
holes in five space-time dimensions,
$Q_1$, $Q_5$, $n$ and $J$ stand respectively for the D1-brane charge,
D5-brane charge, Kaluza-Klein momentum
and the angular momentum (under the $SU(2)_L$ subgroup of the rotation
group).
The third column describes the macroscopic results for the 
logarithmic correction to the entropy under 
the scaling described in the second column. In the last column
a $\surd$ indicates that the microscopic results are available and agree with the
macroscopic prediction while a ? indicates that the microscopic results are
not yet available.} \label{t1} }
\end{table}

Recently Euclidean gravity approach has been used to compute
the logarithmic corrections to the entropy of
a certain class of extremal black holes in string 
theory\cite{1005.3044,1106.0080,1108.3842,1109.0444,1109.3706}. 
Whenever the
corresponding microscopic results are available -- {\it e.g.} for BPS
black hole entropy in four dimensional $\NN=4$ and $\NN=8$
supersymmetric theories, and BMPV black 
hole\cite{9601029,9602065} entropy in
five dimensional string theory -- these macroscopic results are
in perfect agreement with the microscopic 
results\cite{1005.3044,1106.0080,1109.3706}. 
Macroscopic
results are also available for BPS black holes in $\NN=2,3,5$ and $6$
supersymmetric theories in 
four dimensions\cite{1108.3842,1109.0444} but concrete
microscopic results are not yet available. A summary of the current results
on the logarithmic corrections to the entropy of extremal supersymmetric black
holes can be found in \cite{1008.3801}. For the benefit of
the reader we have reproduced this  in table \ref{t1}.
Macroscopic results
also exist for a certain class of extremal non-supersymmetric
black holes\cite{1108.3842,1109.3706,1204.4061}
but there are no microscopic results to compare
them with.

Motivated by this success, in this paper we use the Euclidean gravity
approach to compute logarithmic corrections to the entropy of
non-extremal black holes. This has been done 
before using many different approaches.
Our approach is most closely related to the one due to
Solodukhin, Fursaev and others\cite{9407001,9408068,9412161} 
(see
{\it e.g.} \cite{1104.3712} for a review).
The main difference between our
approach and those reviewed in  \cite{1104.3712} is threefold:
(i) we take into account possible contribution to the conformal anomaly
due to the presence of background fields other than the gravitational
field,  (ii)
we give special treatment to integration over the zero modes and 
(iii) we
keep track of any additional logs which may be generated while
converting the result on the partition function to the result on
entropy via a Laplace transform. 
As our experience with
extremal black holes show, all these effects are important and only after
including these effects we can get agreement between the microscopic
and the macroscopic results. Another technical aspect of our analysis is
that unlike in the approach reviewed in \cite{1104.3712}, 
we do not need to study
quantization of fields in a space-time with conical defect. Instead we
need to compute the partition function of various fields in the euclidean
black hole space-time and interpret the result as the grand canonical
partition function of the black hole.
Although this is not expected to affect the
final result\cite{9407001,9412020}, not having to deal with background
with conical defects 
is particularly important in the context of string theory where the
procedure for quantizing strings in the presence of conical defect
with arbitrary defect angle is not completely understood.

Unfortunately there are no concrete microscopic counting results 
for non-extremal black holes in string theory, and so we cannot at
present use the results of this paper to test string theory. There are
however computations of logarithmic corrections to Schwarzschild
black hole entropy in loop quantum 
gravity\cite{9801080,0002040,0006211,
0401070,0407051,0407052,0411035,0605125,
0905.3168,0907.0846,1006.0634,1103.2723 ,1201.6102}. 
We compare our
result with these results 
and find disagreement. In particular
contribution to the logarithmic correction
from the massless graviton loop seems to be missing from the loop
quantum gravity results. This could be related to the difficulty in
obtaining semiclassical limit in loop quantum gravity. The other case
where microscopic results for non-extremal black holes are available
-- BTZ black holes\cite{0005017} -- the euclidean gravity prediction and the
microscopic results agree trivially\cite{0005003,0104010,0111001}. 
This will be reviewed
in \S\ref{sbtz}.

We shall now summarize our main results. We consider
rotating black hole solution in $D$ dimensional space-time carrying
generic angular momentum so that the symmetry group of the
black hole solution is generated by the Cartan subalgebra of the
rotation group. 
We define the microcanonical entropy $S_{\rm mc}$ to be such
that
$e^{S_{\rm mc}}$ gives the number of quantum states of the black hole
per
unit interval of mass, carrying fixed angular momentum and
charges.\footnote{Throughout this paper we shall work in 
$\hbar=c=G_N=1$ units.} While fixing the angular momentum we fix
only the components along the Cartan generators ({\it e.g.} $J_3$
for the $SO(3)$ rotation group), but sum over all 
values of the Casimirs ({\it e.g.} $\sum_{i=1}^3 J_i J_i$ for $SO(3)$ rotation group).
In this case $S_{\rm mc}$, expressed as a function of the mass
$M$, angular momenta $\vec J$ along the Cartan generators\footnote{Here
the vector sign on $\vec J$ stands not for all components of the
angular momentum but only the Cartan generators. Thus for example
for the SO(3) rotation group $\vec J$ is a one component vector
labelling $J_3$.}
and charges $\vec Q$  receives
a logarithmic correction:
\be \label{eg7int}
S_{\rm mc}(M,\vec J, \vec Q) = S_{\rm BH}(M,\vec J, \vec Q) 
+ \ln a \left( C_{\rm local} 
-{1\over 2} (D-4) -{1\over 2}
(D-2) N_C - {1\over 2} (D-4) n_V\right)\, ,
\ee
where $S_{\rm BH}(M,\vec J, \vec Q)$ is the classical 
Bekenstein-Hawking entropy of a
black holes carrying mass $M$, angular momenta $\vec J$ and charges $\vec Q$,
$a$ is the black hole size parameter, related to the horizon
area $A_H$ via $A_H\sim a^{D-2}$, $N_C=[(D-1)/2]$ is the number of Cartan
generators of the rotation group and
$n_V$ is the number of
$U(1)$ gauge fields. $C_{\rm local}$ is related to the contribution to the
trace anomaly due to the massless fields in the black hole
background\cite{duffobs,christ-duff1,christ-duff2,duffnieu,duffroc,
birrel,gilkey,0306138,1009.4439}. 
In any given theory this can in principle be computed using the procedure
given {\it e.g.} in \cite{0306138}.
 In particular 
$C_{\rm local}$ vanishes in odd dimensions, and for uncharged black holes
in $D=4$ we have (see {\it e.g.} \cite{1104.3712})
\be \label{eclocfinint}
C_{\rm local} = {1\over 90} \, \left( 2 n_S - 26 n_V +7 n_F
- {233\over 2}\, 
n_{3/2} + {424}\right)\, ,
\ee
if the theory contains, besides gravity,
$n_S$ massless scalar fields, $n_V$ 
massless vector fields, $n_F$ massless
Dirac fields and $n_{3/2}$  massless spin 3/2 fields, all
minimally coupled to gravity without any other interactions.
The last term 424 is the contribution from the graviton loop.
In pure gravity theory only this term is present.
In $D=4$ our result \refb{eg7int} differs from the earlier
result given
{\it e.g.} in \cite{9412161} by the $-N_C \ln a=-\ln a$ term that arises
while converting the result on partition function to the result on entropy.
The expression for $C_{\rm local}$ for a Kerr-Newmann black hole in four
space-time dimensions can be computed using  the recent results
on the heat kernel expansion in Einstein-Maxwell theory\cite{1204.4061}
and the result has been given in eq.\refb{eclocfingen}.

We can also consider the entropy in a mixed
ensemble, defined as
\be \label{emixeddef}
e^{S_{\rm mixed}(M, \vec Q)} = \sum_{\vec J}  e^{S_{\rm mc}(M,\vec J, \vec Q)}\, ,
\ee
where  the
sum runs over all eigenvalues of the Cartan generators of the rotation
algebra. Thus $e^{S_{\rm mixed}(M, \vec Q)}\delta M$ 
counts the total number of states
of all angular momentum in the mass range $(M, M+\delta M)$ and fixed values
of the charges, 
with the leading contribution coming from
near
zero angular momentum states where $S_{\rm mc}$ is maximum.
The euclidean gravity analysis leads to the following prediction
for the logarithmic correction to $S_{\rm mixed}$:
\be \label{emixedres}
S_{\rm mixed}(M, \vec Q) = S_{\rm BH}(M, \vec J=0, \vec Q) + \ln a \left( C_{\rm local} 
-{1\over 2} (D-4) - {1\over 2} (D-4) n_V\right)\, .
\ee

Finally we can consider the entropy in the ensemble containing only singlet
states of the rotation group. If we denote the corresponding entropy by
$S_{\rm singlet}$ then $\exp[S_{\rm singlet}(M, \vec Q)]\delta M$ counts the total
number of rotationally invariant 
states in the mass interval $\delta M$ and charge $\vec Q$. Our
result for $S_{\rm singlet}(M, \vec Q)$ takes the form:
\be \label{esinglet}
S_{\rm singlet}(M, \vec Q) = 
S_{\rm BH}(M, \vec J=0, \vec Q)
+ \ln a \left( C_{\rm local} 
-{1\over 2} (D-4) -{1\over 2}
(D-2) N_R - {1\over 2} (D-4) n_V\right)\, ,
\ee
where $N_R=(D-1)(D-2)/2$ is the total 
number of generators of the rotation group.

In 
a theory of pure gravity in $D=4$, \refb{eclocfinint} gives $C_{\rm local}=
212/45$. Also we have $N_R=3$. 
Hence the logarithmic correction to $S_{\rm singlet}$
is given by $\Delta S_{\rm singlet} = \left( {212\over 45}-3\right)\ln a$.
In contrast we find that loop quantum gravity result of
\cite{9801080,0002040,0006211,0401070,0407051,0407052,
0411035,0605125,
0905.3168,0907.0846,1006.0634,1103.2723 ,1201.6102} predicts a result of 
$-2\ln a$.\footnote{For example the 
result of \cite{0002040} gives a logarithmic correction
of $-{3\over 2}\ln A_H
=-3\ln a$ for the entropy of singlet states of the rotation group. 
But this counts the number of states per unit interval in
the area variable $A_H$. Converting this to number of states per
unit interval in mass gives a result of $-2\ln a$.}
The two results obviously disagree.

Since on the macroscopic side the computation is based on one
loop determinant of masless fields in the black hole background,
one could ask whether higher loops can give additional logarithmic
corrections to the black hole entropy. We have checked, based on
naive power counting arguments, that higher loops do not give any
logarithmic corrections to the entropy. This is related to the fact that
in a theory of gravity
infrared divergences become softer at higher loops. It has also
been argued in \cite{1005.3044} that neither massive fields nor
higher derivative corrections to the action can affect the 
logarithmic correction to the entropy. Thus the macroscopic results for
logarithmic corrections seem quite robust.

The rest of the paper is organized as follows. In \S\ref{sgrand}
we describe the computation of logarithmic corrections to the grand
canonical partition functions using Euclidean gravity path integral. In
\S\ref{smicro} we discuss how to translate these results into a statement
of logarithmic corrections to the black hole entropy in different
ensembles.
In \S\ref{sloop} we compare the
macroscopic prediction for the logarithmic correction to the entropy of
a Schwarzschild black hole to the prediction of loop quantum
gravity. 
Finally in \S\ref{sbtz} we review how euclidean gravity approach
can be used to correctly reproduce the logarithmic corrections to
the entropy of a BTZ black hole. Appendix \ref{sa} contains an
analysis of the zero modes of the euclidean black hole solution, and
in appendix \ref{sb} we describe the procedure for removing from the
partition function the contribution due to the thermal gas around the
black hole.

\sectiono{Grand canonical partition function} \label{sgrand}

In this section we shall compute logarithmic corrections to black
hole partition function in 
Einstein's theory of gravity in $D$ dimensions
coupled to a set of massless
abelian vector fields $\{A_\mu^{(\alpha)}\}$, a set of other massless 
neutral scalar fields $\{\vp_s\}$ and also possibly neutral Dirac and
Rarita-Schwinger fields.
We shall assume that the Lagrangian density 
$\LL$ has a scaling property so that
purely bosonic terms all have two derivatives, all terms with two fermion
fields have a single derivative and all terms with four fermion fields
have no derivatives. 
This covers a wide range of theories {\it e.g.} pure
gravity, pure gravity coupled to Maxwell fields, scalars, fermions and other
fields via minimal coupling with no other interaction between these other fields,
a variety of extended supergravity theories  at generic point in the moduli
space of these theories etc. This however excludes theories with cosmological
constant -- we shall comment on them briefly at the end of this section and again
in \S\ref{sbtz} where we discuss the case of BTZ black holes.

\subsection{General framework} \label{s2.1}

Due to the absence of cosmological constant the Minkowski space is
a solution of this theory, and we can consider 
a  charged, rotating
black hole solution which asymptotes to Minkowski space. 
We shall denote by $t$ the time coordinate, by $x^i$ for
$1\le i\le (D-1)$ the spatial coordinates, and by $M$, $Q_\alpha$ and $J_{ij}$
the mass, electric charges and angular momenta
carried by the black hole.\footnote{A black hole can also carry magnetic
charges associated with gauge fields in four dimensions. More generally
in $D$ space-time dimensions a black hole can carry magnetic charges
of $(D-3)$-form gauge fields. 
For simplicity
of notation we shall not explicitly display the dependence of various quantities
on these magnetic charges, but it should be understood that 
one can easily generalize the scaling laws
described in \refb{eg4} if such charges are present.
}
As a consequence of the scaling symmetry mentioned above,
given any classical black hole solution we can generate a whole
family of solutions by a scaling 
\be \label{escale1}
g_{\mu\nu}\to \Lambda^2 g_{\mu\nu}, \quad A^{(\alpha)}_\mu \to 
\Lambda \, A^{(\alpha)}_\mu, \quad \vp_s\to\vp_s\, ,
\ee
and under such a scaling the classical action as well as the
Bekenstein-Hawking entropy scales by $\Lambda^{D-2}$, the mass and
the electric and
magnetic charges scale by $\Lambda^{D-3}$, and the angular momentum
scales by $\Lambda^{D-2}$. 
This leads to the relations
\ben \label{eg4}
&& S_{\rm BH}(\Lambda^{D-3} M,  \Lambda^{D-2} \vec J,
\Lambda^{D-3} \vec Q) = \Lambda^{D-2} 
S_{\rm BH}(M, \vec J,
\vec Q), \nn &&
 a(\Lambda^{D-3} M, \Lambda^{D-2} \vec J,
\Lambda^{D-3} \vec Q) = \Lambda \, a( M, \vec J,
\vec Q)\, ,
\een
where $a$ is the length parameter that gives the size of the black hole,
{\it e.g.} the area of the  event horizon scales as $a^{D-2}$. 

As is well known,
when we analytically continue the black hole solution
to Euclidean space-time,
the time coordinate needs to be periodically identified with period given
by the inverse temperature $\beta$. The other parameters labelling the
Euclidean solution are the
chemical potentials $\mu_\alpha$ dual to the electric charges
$Q_\alpha$ and the
angular velocities $\omega_{ij}$ dual to the angular momenta
$J_{ij}$.  Physically $\mu_\alpha/\beta$ parametrize the component of the
gauge field $A_\mu^{(\alpha)}$ 
along the asymptotic Euclidean time circle, and $\omega_{ij}/\beta$
label the asymptotic values of  $g_{t\phi_{ij}}$ with $\phi_{ij}$ denoting
the angle of rotation in the $x^i$-$x^j$ plane. For the classical black
hole $\beta$, $\mu_\alpha$ and $\omega_{ij}$ are determined in terms
of $M$, $Q_\alpha$ and $J_{ij}$ and vice versa.
Without any loss of generality we can take $\omega_{ij}$ to be
of the form
$\omega_{ij} =\sum_a \omega_a T^a_{ij}$ where $T^a_{ij}$ are the
Cartan generators and $\omega^a$'s are constants.
Thus $\{\omega^a\}$ describe a vector of dimension $N_C$
where $N_C$ is the rank of the rotation group.

In order to calculate the quantum corrections to the black hole entropy
we shall follow the Euclidean path integral
approach since this has successfully reproduced the correct results for
extremal black hole entropy in many 
cases\cite{1005.3044,1106.0080,1109.3706}. 
The euclidean partition function is
defined as\cite{gibbhaw}
\be \label{eg0}
Z(\beta,  \vec \omega, \vec \mu) 
= \int [D\Psi] e^{-\SSS_E(\Psi)}
\ee 
where $\Psi$ stands for all the fields in the theory
including the metric and the gauge
fields, $\SSS_E(\Psi)$ is the Euclidean action 
and the path integral is performed subject to the asymptotic
boundary conditions set by the parameters $\beta$, 
$\vec \omega$
and $\vec\mu$.  While $\beta$ labels the period of the euclidean time
coordinate, $\mu_\alpha/\beta$ denote the component of the 
asymptotic gauge field
$A^{(\alpha)}_\mu$ along the euclidean time, and $\omega^a/\beta$'s 
denote
the $t$-$\phi^a$ components of the asymptotic metric where $\phi^a$ is
the angular coordinate conjugate to the Cartan generator $T^a$. 
Note that in the boundary conditions defining the
path integral there is no reference to a black hole solution; the black
hole becomes relevant as a saddle point which contributes
to the path integral.\footnote{Note also that the Euclidean 
black hole solution is
complex, but this will not affect our analysis since we shall evaluate
the path integral in the saddle point approximation.}
Furthermore, although at the saddle point $M,Q_\alpha$ and $J_{ij}$ are
determined in terms of $\beta$, $\mu_\alpha$ and $\omega_a$, while carrying
out the path integral we only keep fixed $\beta$, $\mu_\alpha$ and $\omega_a$,
and allow fluctuations carrying different values of  $M,Q_\alpha$ and $J_{ij}$.

Classical contribution $Z_{\rm cl}$ to $Z$ is given by 
$\exp[-S_E(\Psi_{\rm cl})]$ where $\Psi_{\rm cl}$ denotes the
classical black hole solution. This is related
to the classical Bekenstein-Hawking entropy 
$S_{\rm BH}(M, \vec J, \vec Q)$ by a Legendre transform\cite{gibbhaw}:
\ben \label{elegend}
&& S_{\rm BH} (M, \vec J, \vec Q)
=  \ln Z_{\rm cl} ( \beta,  \vec \omega, \vec \mu) + \beta \, M
+ \omega^a J_a + \mu_\alpha Q_\alpha, \nn &&
\beta = {\p S_{\rm BH}\over \p M}, 
\quad
\omega_{b} ={\p S_{\rm BH}\over \p J_{b}}, \quad \mu_\alpha
={\p S_{\rm BH}\over \p Q_\alpha}
\nn \Leftrightarrow && 
M= -{\p \ln Z_{\rm cl}\over \p \beta}, 
\quad
J_{b}=-{\p \ln Z_{\rm cl}\over \p \omega_{b} }, \quad Q_\alpha
=-{\p \ln Z_{\rm cl}\over \p \mu_\alpha}
\, , 
\een
where $\{J_a\}$ are related to the angular momenta $\{J_{ij}\}$ via
 $J_{ij} = \sum_a J_a T^a_{ij}$.
{}From \refb{elegend}, \refb{eg4} it follows that
\be \label{ezscaling}
\ln Z_{\rm cl} ( \Lambda \beta,  \vec \omega, \Lambda\vec \mu)
=  \Lambda^{D-2} \ln Z_{\rm cl} ( \beta,  \vec \omega, \vec \mu)\, .
\ee

\subsection{Heat kernel} \label{s2.2}

Our goal in this subsection
is to compute one loop quantum correction to $\ln\, Z$ in the 
limit of  large $\Lambda$ and extract corrections to
$\ln Z$ of order
$\ln\Lambda$ or equivalently $\ln a$. This is done using the heat kernel
technique which we shall now describe.

Let  $\{\phi^\ell\}$ denote the 
set of massless 
fields in the theory. 
Here
the index $\ell$  labels different scalar fields as well as the space-time
indices of tensor fields. Let $\{f_n^{(\ell)}(x)\}$ denote an
orthonormal  basis of eigenfunctions of the kinetic operator
expanded around the near horizon background, with 
eigenvalues
$\{\kappa_n\}$. The orthonormality relations take the form:
\be \label{eortho}
\int d^D x \, \sqrt{\det g} \, G_{\ell\ell'} \, f_n^{(\ell)}(x)\,
f_m^{(\ell')}(x) = \delta_{mn}\, ,
\ee
where $g_{\mu\nu}$ is the metric of the Euclidean black hole
space time and
$G_{\ell\ell'}$ is a metric in the space of fields induced by the
background metric, {\it e.g.} for a vector
field $A_\mu$, $G^{\mu\nu}=g^{\mu\nu}$.
The heat kernel $K^{\ell\ell'}(x,x')$ is defined as
\be \label{eh1}
K^{\ell\ell'}(x,x';s) = \sum_n \, e^{-\kappa_n\, s} \, f_n^{(\ell)}(x)\,
f_n^{(\ell')}(x')\, .
\ee
Among the $f_n^{(\ell)}$'s there may be a special set of
modes for which $\kappa_n$ vanishes.  
We  call these zero modes and define
\be \label{eh1prime}
K^{\prime\ell\ell'}(x,x';s) = {\sum_n}' \, e^{-\kappa_n\, s} \, f_n^{(\ell)}(x)\,
f_n^{(\ell')}(x')\, ,
\ee
where the prime on the sum denotes that we remove the contribution from
the zero modes. 
We also define
\be \label{edefk0}
K(x;s) = G_{\ell\ell'}\, K^{\ell\ell'}(x,x; s), \quad 
K'(x;s) = G_{\ell\ell'}\, K^{\prime\ell\ell'}(x,x; s)\, .
\ee
Using \refb{eortho}-\refb{edefk0} we now get
\be \label{eext23}
 \int d^D x \, \sqrt{\det g}\,   
K'(x;s) = {\sum_{n}}' e^{-\kappa_n\, s}\, ,
\ee
\be \label{eext23pre}
 \int d^D x \, \sqrt{\det g}\,   
K(x;s) =  \int d^D x \, \sqrt{\det g}\,   
K'(x;s) +  N_{\rm zm}\, ,
\ee
where $N_{\rm zm}$ denotes the number of zero modes.

If there are fermion fields present then the definitions of $K$
and $K'$ are modified in two ways. First of all since the fermion
kinetic term is linear in derivatives, we take $\kappa_n$'s to be the
eigenvalues of the square of the fermion kinetic operator and insert
a factor of 1/2 in the trace in \refb{edefk0}. Second 
for fermion fields we insert an extra minus
sign into the trace in \refb{edefk0}.

If we denote by $Z_{\rm nz}$ and $Z_{\rm zm}$ the one loop contribution to the
partition function from integration over the non-zero modes and the zero modes
respectively, then the net result for the partition function to one loop can be expressed
as
\be \label{ezoneloop}
\ln Z = \ln Z_{\rm cl} + \ln Z_{\rm nz} + \ln Z_{\rm zm}\, .
\ee
We shall now discuss the evaluation of $Z_{\rm nz}$ and $Z_{\rm zm}$.

\subsection{One loop contribution to the partition function from the non-zero modes}
\label{snonzero}

The one loop contribution to $\ln Z$ from the non-zero
modes is given by
 \be \label{eloop1pre}
 -{1\over 2}\, {\sum_n}' \ln \kappa_n\, ,
 \ee
 with the understanding that for fermions there is an additional
 factor of $-1/2$ multiplying the summand.
\refb{eloop1pre} of course has many
divergences -- both ultraviolet and infrared -- and to extract something
meaningful we need to understand the role of these divergences. 
First of all we must remember that $Z(\beta,\vec\omega,\vec\mu)$ defined
in \refb{eg0}
describes a grand canonical partition function, and the contribution to
\refb{eg0} from the saddle point corresponding to the 
Euclidean black hole solution can be interpreted as due to a
black hole
in equilibrium with a thermal gas of all the massless (and massive)
particles in the theory.
In the infinite volume limit
the contribution due to the thermal gas
is infinite; so we need to first regularize this by confining the black
hole in some box of size $L$, putting appropriate boundary condition on
all the fields at the boundary of the box. 
The leading contribution to $\ln Z$ from the thermal gas is now
given by $L^{D-1} f(\beta, \vec\omega,\vec \mu)$ where 
$f$ is some function
that scales as $\Lambda^{-D+1}$ under the scaling \refb{eg4},
\refb{ezscaling}. 
There are also possible subleading contributions involving lower positive
powers of $L$ due to boundary effects.
We
must subtract all these contributions from $\ln Z$ in order to identify the
contribution to the partition function associated with the black hole
microstates. To do this we introduce a length $a_0$ which is fixed
but large
compared to the Planck length,
consider another black hole solution
which is related to the original solution by a rescaling of the parameters
described in \refb{eg4},
\refb{ezscaling} with $\Lambda=a_0/a$ and confine this new
system in a box of side
$L_0=L a_0/a$. 
In the common coordinate
system in which the metric for the 
new and the original black hole solutions are
simply related by a multiplicative factor of $(a_0/a)^2$, the
shape of the box in which we confine the two black holes  
are taken to be identical, and furthermore the boundary conditions
on the various fields are taken to be related by the scale transformation
\refb{escale1} with $\Lambda=a_0/a$.
Let $Z_0$ be
the partition function of the new black hole solution.
The leading contribution to $\ln Z_0$
from the thermal gas is given by $(L a_0 /a)^{D-1}
f(a_0 \beta/a,  \vec \omega, a_0\vec \mu/a)
= L^{D-1} f(\beta, \vec\omega,\vec \mu)$. This is identical to the leading
thermal gas contribution to $\ln Z$ and hence subtracting $\ln Z_0$
from $\ln Z$ has the effect of removing the leading
contribution to $\ln Z$ due to the thermal gas.
In fact we have argued in appendix \ref{sb} that subtracting
$\ln Z_0$ from $\ln Z$ also removes the spurious boundary terms,
proportional to subleading powers of $L$,
which may be present.
On the other hand since the new black hole has a fixed size $a_0$,
subtracting $\ln Z_0$ does not remove the $a$ dependent contribution
to $\ln Z$ that comes from the intrinsic entropy of the black hole of
size $a$.
By an abuse of notation we shall continue to denote the regulated
partition function $Z/Z_0$
by the same symbol $Z$.
If we denote by $\kappa^{(0)}_n$ the eigenvalues of the
kinetic operator of the new black hole solution  then 
the one loop contribution to $\ln Z$ from the non-zero modes
after the subtraction is
given by
 \be \label{eloop1}
 \ln Z_{\rm nz}=
 -{1\over 2}\, {\sum_n}' (\ln \kappa_n - \ln \kappa^{(0)}_n)\, .
 \ee
 
We shall compute \refb{eloop1} using 
Schwinger proper time formalism.
We use the relation
\be \label{ereln}
\lim_{\eps\to 0} \int_\eps^\infty {ds\over s}\, 
\left(e^{-As} - e^{-Bs}\right) =
\ln{B\over A}\, ,
\ee
to express \refb{eloop1} as
\be \label{eloop3old}
\ln Z_{\rm nz}={1\over 2}
\int_\eps^\infty {ds\over s} {\sum_n}'
(e^{-\kappa_n s} - e^{-\kappa^{(0)}_n s})\, ,
\ee
where $\eps$ is an $a$-independent 
ultraviolet cut-off.
In an ultraviolet regulated theory $\eps$ is of the order of
Planck length$^2$ which in our convention is of order unity.
The final result will not depend on the details of this cut-off.

To proceed further we note that the two black hole solutions
of size $a$ and $a_0$ as well as their infrared cut-offs $L$ and
$L a_0/a$ are  related by a rescaling of the metric and the
gauge fields as given in \refb{escale1} with $\Lambda=a_0/a$.
It follows from this that the eigenvalues $\kappa_n$ and $\kappa^{(0)}_n$
are related as
\be \label{ekappan0}
\kappa_n^{(0)} =\kappa_n a^2 / a_0^2\, .
\ee
Hence \refb{eloop3old} can be written as
\be \label{eloop3}
\ln Z_{\rm nz}={1\over 2}
\int_\eps^\infty {ds\over s} {\sum_n}'
e^{-\kappa_n s} - {1\over 2}
\int_\eps^\infty {ds\over s} {\sum_n}'
e^{-\kappa_n sa^2 / a_0^2} = 
{1\over 2}
\int_\eps^{\eps a^2 / a_0^2}  {ds\over s} {\sum_n}'
e^{-\kappa_n s} \, ,
\ee
where in the last step we have carried out a rescaling of variable
$sa^2 / a_0^2\to s$ in the second term.
Using \refb{eext23}, \refb{eext23pre} 
we can express \refb{eloop3} as\be \label{e4}
\ln Z_{\rm nz}
=
{1\over 2}\, \int_{\eps}^{\eps a^2 / a_0^2}\, {ds\over s} \, 
\left(  \int d^D x \, \sqrt{\det g}\,   K(x;s) - N_{\rm zm}\right)\, .
\ee
{}From \refb{e4} we see that the variable $s/a^2$ ranges between
$\eps/a^2$ and $\eps/a_0^2$, and hence remains small over the
entire integration range since both $a$ and $a_0$ are taken to be
large compared to the length cut-off $\sqrt\eps$..
This allows us to use the small $s$ expansion of $K(x;s)$.
In $D$ dimensions $K(x;s)$ has a small $s$ expansion
of the form (see {\it e.g.} \cite{0306138})
\be \label{expansion}
K(x;s) =\sum_{n=0}^\infty \, K_{-{D\over 2}+n}(x) \, s^{-{D\over 2}+n}\, ,
\ee
where the coefficients $K_{-{D\over 2}+n}(x)$ are given by local 
general coordinate and gauge invariant combinations of the background
fields containing $2n$ derivatives, {\it e.g.} $K_{-D/2}(x)$ is a
constant, $K_{-D/2 +1}$ is a linear combinations of
$R$, $F^{(\alpha)}_{\mu\nu} F^{(\alpha')\mu\nu}$, 
$\p_\mu\vp_s \p^\mu \vp_{s'}$ etc.
The logarithmic correction comes from the order $s^0$ terms in this
expansion.
Using \refb{e4} and \refb{expansion} we get
\be \label{enetlognz}
\ln Z_{\rm nz}
=
\ln {a}\, \left(  C_{\rm local} - N_{\rm zm}\right) +\cdots
\, ,
\ee
where\be \label{eclocal}
C_{\rm local} =  \int d^D x\, \sqrt{\det g}\,   K_0(x) \, ,
\ee
and $\cdots$ denotes other terms which do not have $\ln a$ factors but
are possibly divergent in the $\eps\to 0$ limit. The significance of these
divergent terms will be explained shortly.
\refb{expansion} shows that $K_0(x)$ and hence
$C_{\rm local}$ vanishes in odd dimensions.
In even dimensions $K_0(x)$ contains $D$ derivatives
and as a result $C_{\rm local}$ is a function of the variables
$(\beta,\vec \omega,\vec\mu)$ which scales as $\Lambda^0$
under the scaling \refb{eg4}.

Before we proceed, a few comments are in order.
\begin{enumerate}
\item 
Special mention must be made of the terms in
\refb{expansion} which diverge in the $\eps\to 0$ limit.
The interpretation of these divergences was discussed in 
\cite{9408068}.
A term of the form
$K_{-\alpha}(x) s^{-\alpha}$ gives a contribution
\be \label{econtri1}
{1\over 2}\, \int d^D x\, \sqrt{\det g}\, 
\int_{\eps}^{\eps a^2/ a_0^2}\, {ds\over s} \, K_{-\alpha}(x)
s^{-\alpha}
= {1\over 2}\, \int d^D x\, \sqrt{\det g}\, K_{-\alpha}(x)\, {1\over \alpha} \, 
\eps^{-\alpha} ( 1 - (a/a_0)^{-2\alpha})\, . 
\ee
Now since $K_{-\alpha}$ contains $D/2 -\alpha$ derivatives,
it follows from 
the scaling symmetry of the theory that $\int d^Dx \, \sqrt{\det g}\,
K_{-\alpha}(x)$ has $a$ dependence of the form $a^{2\alpha}$.
Thus \refb{econtri1} goes as $a^{2\alpha}/\eps^\alpha
- a_0^{2\alpha}/\eps^\alpha$ and is
divergent  for $\alpha> 0$ if we insist on taking the $\eps\to 0$ limit
instead of keeping $\eps$ of the order of Planck length square.
What is the origin of these
divergences? 
Clearly the $a^{2\alpha}$ term comes from the original black hole
solution and the $a_0^{2\alpha}$ term comes from the black hole
of size $a_0$; so it is sufficient to focus on the $a^{2\alpha}$ terms.
The divergent coefficient of this term in the $\eps\to 0$ limit
can be traced to the usual ultraviolet divergences
in field theory which renormalize various parameters of the theory.
For example the leading divergent
term, corresponding to $\alpha=D/2$, has $K_{-\alpha}(x)$ an
$x$ independent constant. The corresponding divergent contribution to
\refb{econtri1} can be written as
\be \label{econtri2}
{1\over 2}\, {K_{-D/2}\over D/2}\,  \eps^{-D/2}\, 
\int d^D x\, \sqrt{\det g}
\, .
\ee
\refb{econtri2} 
clearly has the interpretation of a one loop
contribution
to the Euclidean effective action of the form 
$-{1\over 2}\, {K_{-D/2}\over D/2}\,  \eps^{-D/2}\, 
\int d^D x\, \sqrt{\det g}$ -- a cosmological constant term. Since the
theory we consider by assumption does not have a cosmological constant
this must be removed by a counterterm. The same counterterm will also
remove the corresponding divergent
contribution to \refb{e4}. Similarly
the first subleading divergent contributions proportional to
$\eps^{-D/2+1} K_{-D/2+1}(x)$ can be interpreted as the result of
renormalization of the coefficients of various two derivative terms in the
action, {\it e.g.} $R$, $F^{(\alpha)}_{\mu\nu} F^{(\alpha')\mu\nu}$, 
$\p_\mu\vp_s \p^\mu \vp_{s'}$ etc. Again these divergences must be
removed by adding counterterms to the action, and these will have the
effect of removing the corresponding divergences from 
\refb{e4}.\footnote{If the theory contains equal number of bosonic and fermionic
degrees of freedom, {\it e.g.} a supersymmetric theory, then there is no one
loop contribution to the renormalization of the cosmological constant and
$K_{-D/2}$ vanishes. In
some extended supergravity theories the one loop correction to two derivative
terms also vanish. In  this case $K_{-D/2+1}$ is also zero.}
This way all the divergent contributions to $\ln Z$ are removed by
adding to the action the same
local counterterms which are needed to
get finite results for physical quantities independently of the
computation of black hole entropy. Alternatively if the theory comes with an
intrinsic ultraviolet cut-off that makes $\eps$ of the order of Planck length
square, then these contributions can be absorbed into a finite renormalization
of the various coupling constants of the theory.
\item
Since we have put the
back hole inside a box of size $L$ with appropriate boundary conditions
on various fields at the boundary of the box, the small $s$ expansion
of $K(x,s)$ also contains
terms which are localized on the boundary instead of the
bulk of space-time\cite{0306138}. 
These contributions to the partition function
do not have anything to do with the black hole, and arise from
boundary effects. Thus they
must be removed from $\ln Z$. Formally this will be done by explicitly
removing from $K(x;s)$ the boundary terms carrying
non-positive powers of $s$ in the small $s$ expansion. 
This significance of this subtraction will be
explained in appendix \ref{sb}. The positive powers of $s$
on the other hand will give negligible contribution to \refb{e4}.
\item
Eq.\refb{e4}, 
\refb{expansion} 
shows that in general the one loop
correction to $\ln Z$ and hence
the entropy depends on the form of the black hole solution
over the entire space-time, and not just the near horizon geometry. This
would seem to be in apparent conflict with computations based on Wald's
formula applied to the quantum effective action
or entanglement entropy computation reviewed in 
\cite{1104.3712}, which depend only on the near horizon
geometry. We must however keep in mind that the near horizon geometry
of the black hole, expressed as a function of the asymptotic parameters
like temperature, chemical potential etc., can get corrected due to quantum
corrections to the effective action
and these corrections are controlled by the form of the original
solution over the entire space-time. In particular the corrections to the near
horizon geometry could
involve terms proportional to $\ln a$ and hence the usual Bekenstein-Hawking
entropy evaluated in the new background could have additional
logarithmic corrections. Thus in order to compute the
logarithmic correction to the entropy we must use information about the full
black hole solution in all the approaches. Extremal
black holes are exceptional since for them
the attrator mechanism allows us to fix
the near horizon geometry
without knowing the details of the full solution. Also for uncharged black holes 
in $D=4$
the analysis simplifies since $K_0(x)$ is just the Euler density and hence
its integral is determined by the topology of the solution. At the same time
it does not affect the equations of motion and hence does not introduce
any correction proportional to $\ln a$ to the near horizon field configurations.
\item
It is instructive to identify the region of loop momentum integration that is
responsible for the $\ln a$ terms in the partition function. For this we shall take
the unltraviolet cut-off $\eps$ to be of order unity, \i.e.\ of the order of
the square of the Planck length. Then the $\ln a$ contribution to
\refb{e4} comes from the range $1<< s << a^2/a_0^2$. In terms of loop
momentum -- which is of order $1/\sqrt s$ -- 
this means that the logarithmic corrections come from the
range of loop momentum integration which is much less than the Planck
mass. Hence
it involves infrared physics.
\end{enumerate}

Let us now briefly discuss the evaluation of $K_0(x)$. 
It follows from \refb{expansion} that
$K_0(x)$
vanishes for odd $D$.
In even dimensions we can evaluate
$K_0(x)$ in a general black hole background using the general
method described in \cite{0306138}, which allows us to express
$K_0(x)$ as a linear combinations of covariant terms, each containing
$D$ derivatives. Thus for example in $D=4$, $K_0(x)$ will have
the form
\be \label{ekoxgen}
K_0(x) = \alpha \, R_{\mu\nu\rho\sigma} R^{\mu\nu\rho\sigma} 
+ \beta R_{\mu\nu} R^{\mu \nu} + \gamma R_{\mu\nu\rho\sigma}
F^{\mu\nu} F^{\rho\sigma} 
+ \cdots \, ,
\ee
where $\alpha$, $\beta$, $\gamma$ etc. are computable coefficients.
For Einstein-Maxwell theory in four space-time dimensions this has been
calculated recently in \cite{1204.4061} with the result:
\be \label{ea4fin}
K_0(x) = {1\over 360 \times 16\pi^2} 
\left(398 R_{\mu\nu\rho\sigma} R^{\mu\nu\rho\sigma}
+ 52 R_{\mu\nu} R^{\mu\nu}\right)\, .
\ee
If in addition we have $n_S$ scalars, $n_F$ Dirac fermions,
$(n_V-1)$ more vector fields and $n_{3/2}$ spin 3/2 fields, 
all minimally coupled to background gravity
and no coupling to the background gauge field, then
\refb{ea4fin} is modified to\cite{duffobs,christ-duff1,christ-duff2,duffnieu,duffroc,
birrel,gilkey,0306138,1009.4439}
\ben \label{ea4finmore}
K_0(x) &=& {1\over 360 \times 16\pi^2} 
\Bigg\{\left(398+2n_S-26 (n_V-1)+ 7 n_F-{233\over 2} n_{3/2}\right)
 R_{\mu\nu\rho\sigma} R^{\mu\nu\rho\sigma}
\nn
&& \qquad \qquad \qquad
+ \left(52-2 n_S+ 176 (n_V-1)+ 8 n_F+ 233 n_{3/2}
\right) R_{\mu\nu} R^{\mu\nu}\Bigg\}\, .
\een

As an illustration we shall now compute $K_0(x)$ for a Kerr-Newmann
black hole.
The metric of a 
general Kerr-Newmann black hole of mass $M$, angular momentum
$J$ and charge $Q$ (in appropriate units) is given by
\ben \label{ekerrnewmetric}
ds^2 &=& - {r^2 + \aaa^2 \cos^2 \psi - 2 Mr +Q^2\over r^2 + \aaa^2 \cos^2 \psi} dt^2
+ { r^2 + \aaa^2 \cos^2 \psi   \over   r^2 + \aaa^2  - 2 Mr +Q^2} dr^2 
+(r^2 + \aaa^2 \cos^2 \psi) d\psi^2 \nn &&
+
{ (r^2 + \aaa^2 \cos^2 \psi) (r^2+\aaa^2) + (2 Mr -Q^2) \aaa^2\sin^2\psi
\over  r^2 + \aaa^2 \cos^2 \psi} \sin^2\psi d\phi^2 
\nn &&
+{2 (Q^2-2 M r) \aaa\over r^2 + \aaa^2 \cos^2 \psi} \sin^2\psi \, dt d\phi \nn 
\aaa &=& {J\over M}\, .
\een
The location $r_H$ of the horizon and the classical Bekenstein-Hawking entropy
$S_{BH}$ are given respectively by
\be \label{ehorlockn}
r_H = M + \sqrt{M^2 - Q^2 - \aaa^2} =
{1\over M} (M^2 + \sqrt{M^4 - Q^2M^2 - J^2}) \, ,
\ee
and
\be \label{esentropy}
S_{BH} = \pi (2M^2 - Q^2 +2M\sqrt{M^2 - (\aaa^2 + Q^2)})
= \pi (2M^2 - Q^2 +2\sqrt{M^4 - (J^2 + Q^2M^2)})\, .
\ee
Using \refb{elegend} we now get
\ben \label{entropykn}
\beta &=& {2\pi M\over \sqrt{M^4 - J^2 - M^2 Q^2}}
\, \bigg\{ 2M^2 - Q^2 + 2\sqrt{M^4 - J^2 - M^2 Q^2}\bigg\}\, ,\nn
\omega &=& - {2\pi J\over \sqrt{M^4 - J^2 - M^2 Q^2}}\, , \nn
\mu &=& -{2\pi Q\over \sqrt{M^4 - J^2 - M^2 Q^2}} \left\{
M^2 + \sqrt{M^4 - J^2 - M^2 Q^2}\right\}\, .\een

Now for the background \refb{ekerrnewmetric} we have
\cite{9912320,0302095}
\ben \label{erimric}
R_{\mu\nu\rho\sigma}R^{\mu\nu\rho\sigma}
&=& {8\over (r^2+\aaa^2 \cos^2\psi)^6}
\Big\{ 6 M^2 (r^6 - 15\aaa^2 r^4\cos^2\psi
+ 15 \aaa^4 r^2 \cos^4\psi - \aaa^6 \cos^6\psi) \nn
&& -12 M Q^2 r (r^4 - 10 r^2 \aaa^2 \cos^2\psi + 5 \aaa^4 \cos^4\psi) \nn &&
+ Q^4 (7r^4 - 34 r^2 \aaa^2 \cos^2\psi + 7 \aaa^4 \cos^4\psi)
\Big\} \nn
R_{\mu\nu} R^{\mu\nu} &=& {4 Q^4 \over (r^2 + \aaa^2 \cos^2 \psi)^4} \, , \nn
\det g &=& (r^2+\aaa^2 \cos^2\psi)^2 \, \sin^2\psi\, .
\een
After analytic continuation $t\to -i\tau$ and identifying $\tau$
as a periodic variable with period $\beta$ we get,
\ben \label{eintegral}
&& \int d^4 x \sqrt{\det g} \, R_{\mu\nu} R^{\mu\nu} \nn
&=& \frac{\pi \beta Q^4}{2 \aaa^5 r_H^4 \left(\aaa^2+r_H^2\right)} 
\left\{3 \aaa^5 r_H+2 \aaa^3 r_H^3+3 \left(\aaa^2-r_H^2\right) 
\left(\aaa^2+r_H^2\right)^2 \tan
   ^{-1}\left(\frac{\aaa}{r_H}\right)+3 \aaa r_H^5\right\}
   \nn
&& \int d^4 x \sqrt{\det g} R_{\mu\nu\rho\sigma} R^{\mu\nu\rho\sigma} 
 = 64\pi^2 + 4\int d^4 x \sqrt{\det g} R_{\mu\nu} R^{\mu\nu}
\een
Substituting these into eqs.\refb{eclocal}, \refb{ea4finmore}  we get
\ben \label{eclocfingen}
C_{\rm local} 
&=&  {1\over 90} \, \left( 2 n_S - 26 (n_V-1) +7 n_F
- {233\over 2}\, 
n_{3/2} + {398}\right) \nn && 
+ {(1644 + 6 n_S + 72 (n_V-1) + 36 n_F - 233 n_{3/2} )\over 360 \times 16\pi^2}
 \frac{\pi \beta Q^4}{2 \aaa^5 r_H^4 \left(\aaa^2+r_H^2\right)} \nn && \times
\left\{3 \aaa^5 r_H+2 \aaa^3 r_H^3+3 \left(\aaa^2-r_H^2\right) 
\left(\aaa^2+r_H^2\right)^2 \tan
   ^{-1}\left(\frac{\aaa}{r_H}\right)+3 \aaa r_H^5\right\}
 \, .
\een

We can now consider some special cases of this formula:
\begin{enumerate}
\item For an uncharged black hole we have $Q=0$ and \refb{eclocfingen}
reduces to
\be \label{eclocfin}
C_{\rm local} = {1\over 90} \, \left( 2 n_S - 26 (n_V-1) +7 n_F
- {233\over 2}\, 
n_{3/2} + {398}\right)\, .
\ee
Since this is valid for all angular momentum, it is also valid
for Schwarzschild black holes. This agrees with the standard results
in the literature, see {\it e.g.} \cite{9412161}.
\item For charged
non-rotating black hole we have $\aaa=0$. Taking the $\aaa\to 0$ limit
of \refb{eclocfingen} we get 
\ben \label{ecloccharged}
C_{\rm local} &=& {1\over 90} \, \left( 2 n_S - 26 (n_V-1) +7 n_F
- {233\over 2}\, 
n_{3/2} + {398}\right) \nn &&
+ {\beta Q^4 \over 1800\pi  r_H^5} 
{(1644 + 6 n_S + 72 (n_V-1) + 36 n_F - 233 n_{3/2} )}
  \, .
\een
\end{enumerate}

\subsection{One loop contribution to the partition function from the
zero modes} \label{szeromodes}

Next we turn to the evaluation of the number $N_{\rm zm}$ 
of zero modes and their contribution $Z_{\rm zm}$ to the partition 
function. 
The zero
modes of the Lorentzian black hole solution
are associated
with the translation and rotation symmetries which are broken by
the black hole. It has been argued in appendix \ref{sa}
that only those translations which are invariant under the rotation
generator $\vec \omega\cdot \vec T$ generate zero modes of the Euclidean
black hole solution, -- the other translational zero modes
and all the rotational
zero modes of the Lorentzian solution fail to satisfy the required
periodicity along the Euclidean time direction and hence are
lifted.
We shall denote by $n_T$ the number of  translational zero modes
of the Euclidean solution.
Thus for example in four space-time dimensions $n_T=3$ for non-rotating
black holes since for these $\vec\omega\cdot \vec T=0$ and hence all
three broken translation symmetries generate zero modes. But for a 
black hole rotating along the $z$-axis we have $\vec\omega\cdot \vec T=T_3$
and hence only the translation along the third direction generates a zero mode.
Thus in this case we have $n_T=1$.
A rotating black hole also breaks part of the rotational
invariance, but as mentioned above there are no
zero modes of the Euclidean black hole solution associated with the
broken rotational invariance.

Let $h_{\mu\nu}$ denote the fluctuating gravitons
in the black hole background. We normalize the path integral
measure as
\be \label{ebp2bmp}
\int [Dh_{\mu\nu}] \exp\left[- \int d^{D} x \, \sqrt{\det g} \, 
g^{\mu\nu} g^{\rho\sigma} h_{\mu\rho} h_{\nu\sigma}
\right] = 1\, .
\ee
Now in an appropriate coordinate system
the metric $g_{\mu\nu}$ has the form $a^2 g^{(0)}_{\mu\nu}$ where
$g^{(0)}_{\mu\nu}$ is $a$ independent. 
Then we can express \refb{ebp2bmp} as
\be \label{ebp3bmp}
\int [Dh_{\mu\nu}] \exp\left[- a^{D-4} \int d^{D} x \, 
\sqrt{\det g^{(0)}} \, 
g^{(0)\mu\nu} g^{(0)\rho\sigma}h_{\mu\rho} h_{\nu\sigma}
\right] = 1\, .
\ee
Thus the correctly normalized integration
measure, up to an $a$ independent constant, is 
$\prod_{x,(\mu\nu)} d(a^{(D-4)/2}h_{\mu\nu}(x))$.
Now the translational zero modes are associated
with diffeomorphisms with non-normalizable parameters
$\xi^{(i)\mu}$ such that $\xi^{(i)\mu}\to \delta_{i\mu}$
asymptotically but vanishes below a certain radius. 
We can introduce parameters $u_{(i)}$ labelling
the zero mode deformations via
\be \label{etrs1}
h_{\mu\nu} = u_{(i)} \left(D_\mu\xi^{(i)}_\nu + D_\nu\xi^{(i)}_\mu\right)\, .
\ee
Our strategy will be to first transform integration over the
metric variables to integration over $u_{(i)}$ 
and then find the $a$ dependence
of the range of integration over the $u_{(i)}$'s.
$D_\mu\xi^{(i)}_\nu + D_\nu\xi^{(i)}_\mu$ is the Jacobian of
change of variables from $h_{\mu\nu}$ to $u_{(i)}$. 
Although with our choice of normalization $\xi^{(i)\mu}$ is $a$
independent, lowering the index makes $\xi^{(i)}_\mu\sim a^2$.
Thus for each zero mode  the Jacobian of change of variables
from $a^{(D-4)/2} h_{\mu\nu}$ to $u_{(i)}$ gives a factor of
$a^{{D-4\over 2} + 2}=a^{D/2}$. Next we need to
find the $a$ dependence of the integration range over the
$u_{(i)}$'s. Since $\xi^{(i)\mu}\to \delta_{i\mu}$ asymptotically,
the integration range of $u_{(i)}$ corresponds to the range of the
coordinate $x^i$ in which we confine the black hole. Let $L$ be the
proper size of the box in which we confine the black hole -- for simplicity
we shall take this to be the same in all directions.
Now in the coordinate
system in which the metric has the form $a^2 g^{(0)}_{\mu\nu}$,
the asymptotic metric is $a^2 \eta_{\mu\nu}$. 
Thus the range of the coordinate $x^i$ is given by $L/a$, and hence
the integration range over $u_{(i)}$ is also given by $L/a$. Combining this
with the Jacobian factor $a^{D/2}$ found earlier we see that integration over
each $u_{(i)}$ produces a factor of $a^{D/2} (L/a)$. 
Thus the
net contribution from $n_T$ such zero modes to 
$\ln Z$ is given by
\be \label{etrntcon}
\ln Z_{\rm zm}= {D\over 2} n_T \, \ln a + n_T \ln {L\over a}\, .
\ee
Using \refb{enetlognz},
\refb{etrntcon} and $N_{\rm zm}=n_T$ 
we can express \refb{ezoneloop} as
\be \label{egrfin}
\ln Z = \ln Z_{\rm cl}+ n_T \ln L+ 
\left( C_{\rm local}  + {1\over 2} n_T (D-4)\right) \, \ln a +\cdots\, .
\ee

\subsection{Higher loop contributions} \label{s2.3}

Since we have only analyzed the contribution from one loop
determinants of massless fields it is appropriate to ask if higher
loop corrections could change the result. 
In $D$ dimensions
naive power counting shows that the $\ell$-loop vacuum graph has
a net mass dimension of $(D-2)\ell + 2$, so that after multiplying this
by $(\ell-1)$ powers of
Newton's constant $l_p^{D-2}$ we get a term of mass
dimension $D$ -- the required dimension of a Lagrangian
density. 
The contribution to
$\ln Z$ is obtained by multiplying this by a factor of $\sqrt{\det g}
\sim a^D$ and then integrating this over the 
black hole space-time. 
Furthermore in the presence of the black hole background the
various propagators and vertices carrying momenta $k$
are modified from their
form in flat space-time background by multiplicative functions
of $ka$ which approach 1 for large values of $ka$,
so
that for large momentum we recover the propagators and
vertices in flat space-time background. 
Putting these results together we see that the $\ell$-loop
contribution to $\ln Z$ may be schematically written as
\be \label{elp1}
l_p^{(D-2)(\ell-1)} \, a^{D} \, \int^{1/\sqrt \eps} d^{D\ell} k \, k^{2-2\ell} \, F(ka)\, ,
\ee
where $F(ka)$ is some function that approaches 1 for large value of
its argument, and the factor of $a^D$ outside the integral comes from the
$\sqrt {\det g}$ factor. The power of $k$ inside the integral has been adjusted
so that $d^{D\ell} k \, k^{2-2\ell}$ has mass dimension
$(D-2)\ell + 2$. The upper limit $1/\sqrt\eps$ on the integral indicates that
the ultraviolet cut-off on the loop momentum integral 
is taken to be of order $1/\sqrt\eps$. 
Using a change of variables $\wt k = ka$ we can express \refb{elp1} as
\be \label{elp2}
l_p^{(D-2)(\ell-1)} \, a^{-(D-2)(\ell-1)} \, \int^{a/\sqrt \eps} d^{D\ell} \wt k \, \wt 
k^{2-2\ell} \, F(\wt k)\, .
\ee

First consider the case where all loop momenta are of the same order.
Since $F(\wt k)\to 1$ for large $\wt k$, we can expand it in a power
series in $\wt k^{-1}$. Possible $\ln a$ term will come from the
order $\wt k^{-(D-2)\ell -2}$ term in this expansion. But the
corresponding $\ln (a/\sqrt \eps)$ term is multiplied by
$a^{-(D-2)(\ell-1)}$ and hence is suppressed for large $a$ except
for $\ell=1$. This shows that we do not get any logarithmic correction
to the entropy from the region of loop momentum integration where
all momenta are of the same order. Next consider the possibility where a
subset of the loop momenta are smaller than the rest; we shall call the
part of the graph that carries low momentum soft part and the rest hard
part. In that case we can regard the effect of the hard loops as 
renormalization of
the vertices and propagators of the soft part of the graph,
and the contribution from such graphs essentially reduces to a lower order
contribution where soft lines appear as propagators and all hard lines are
collapsed into renormalization of vertices and propagators.
Thus 
as long as this
renormalization does not change the low energy effective
action {\it e.g.} the massless
particles are kept massless 
and minimal coupling to gravity remains minimal
(by adding explicit counterterms if necessary) these contributions do not
change logarithmic corrections to the black hole entropy. 
This argument holds in 
particular in a theory of pure gravity, since 
gravitons must remain massless
even after quantum corrections if they are to describe the long range 
gravitational force that we see in nature.
The renormalization effects can also generate higher derivative couplings,
but higher derivative corrections will give additional powers of
$l_p/a$, making the coefficient of the $\ln (a/\sqrt \eps)$ term
suppressed by even more powers of $l_p/a$ than what has been
argued before.
Thus we conclude
that as long as 
the massless fields are kept massless and minimally coupled to gravity
even after renormalization effects are taken into account, the one
loop logarithmic correction to the partition function is not altered by
higher loop corrections.

\subsection{Effect of cosmological constant term} \label{s2.4}

Finally let us discuss briefly how the analysis changes
in the presence of a cosmological constant term. In this case the
eigenvalues of the kinetic operator will be of order $l^{-2}$ where
$l$ is some length scale 
set by the cosmogical constant, and the presence of a black hole 
of size $a$ introduces
small corrections of order $a^{-2}$ to these eigenvalues. 
Thus the integration over the proper time variable
$s$ will be suppressed exponentially for $s>>l^{2}$ and there is no
contribution of order $\ln a$ from the region $l_p^2<< s << a^2$. 
As a result there is no logarithmic correction to $\ln Z$ from the
non-zero modes. Depending on the situations there may be some
logarithmic correction from the zero modes, but often, as in the case
of BTZ black holes to be discussed in \S\ref{sbtz}, even the zero mode
contributions are absent. In such cases there is no correction of order
$\ln a$ to $\ln Z$. However as we shall see in \S\ref{smicro} the
entropy can still receive logarithmic corrections during
the process of converting the partition function into 
entropy.\footnote{The effect of the change in the ensemble
on the logarithmic corrections to the entropy has been analyzed before,
see {\it e.g.} 
\cite{0111001,0205164,0309026,0402173,1108.4670}.}

\sectiono{Black hole entropy in microcanonical and other ensembles}
\label{smicro}

The $Z$ computed in \S\ref{sgrand} describes the partition
function computed from the gravity side.
This should be identified as
the statistical grand canonical partition
function $Z_{stat}$ 
\be \label{egrdef}
Z_{stat} = Tr \left( e^{-\beta E  - \vec \omega.\vec J
- \vec \mu\cdot \vec Q}\right)\, ,
\ee
where the trace runs over all the black hole microstates carrying 
different mass, charges, angular momenta and
momenta
$\vec P$.  $E=M + \vec P^2/2M$ denotes the total energy of the
black hole. $Z_{stat}$ can be
computed from microstate degeneracies whenever the latter
results are available, which can then be compared with the
partition function $Z$ computed from the gravity side.
Alternatively, by equating $Z_{stat}$ with the Euclidean gravity 
prediction \refb{egrfin}
for $Z$ we can arrive at
definite predictions for the black hole entropy in the
microcanonical (or, in any other) ensemble. This can 
then be
compared with the microscopic results for the same 
quantities if and when the microscopic results are
available.
We shall follow the latter point of view and, from now on, identify
$Z_{stat}$ with $Z$.
Our goal in this section will be to convert the
result for $Z$ obtained in 
\S\ref{sgrand} to entropies defined in different ensembles
which may be relevant for comparison with the microscopic 
results.

\subsection{Entropy in the microcanonical ensemble} \label{smican}

Before we proceed we need to clarify the meaning of
the entropy in 
microcanonical ensemble. Since the charges  are
quantized in integer units, 
it is  possible to fix their values at definite numbers
and count states to define the entropy. For angular momentum we cannot
fix all the components, but we can fix the 
components associated with the Cartan generators to
specific integer or half integer values.
This cannot be done for the mass
since the quantization rules for the mass is not known without knowing the
microscopic details. For this reason we define the microcanonical entropy
such that
\be \label{edefmicro}
e^{S_{\rm mc}(M, \vec J, \vec Q)} \, \delta M\, 
\ee
represents the number of internal 
microstates of the black hole
in the mass range $\delta M$, carrying charge $\vec Q$, angular momentum
$\vec J$ and vanishing total momentum.\footnote{In contrast the entropy
of extremal black holes analyzed in 
\cite{1005.3044,1106.0080,1108.3842,1109.0444,1109.3706,1204.4061} 
correspond to ground state
degeneracy in a given charge sector, and not the number of states in a
given range of mass. Thus we cannot directly compare our results with
those of \cite{1005.3044,1106.0080,1108.3842,1109.0444,1109.3706,1204.4061}.
In any case since the present analysis has been carried out under the
assumption that all length scales are of the same order -- in particular
$\beta \sim a$ -- the results are not necessarily valid in the extremal limit
in which $\beta\to\infty$ keeping $a$ fixed.}
The interval $\delta M$
apearing in \refb{edefmicro} must be small enough so 
that
$S_{\rm mc}$ does not vary appreciably over the interval, and yet large
enough so as to contain a large number of states and so as to be
larger than the decay width of individual microstates. This can be achieved by
taking the size of the black hole to be sufficiently large.

Since $S_{\rm mc}$ does not depend on the momentum $\vec P$ as a
consequence of Lorentz invariance, we can perform the sum over
$\vec P$ implicit in \refb{egrdef} explicitly.
If $\vec P$ is not
invariant under $\vec \omega.\vec J$, then $e^{-\vec\omega.\vec J}$ acting on
the state will produce a state with a different momentum\footnote{For
this we analytically continue $\omega_a$ to imaginary values so that
$e^{-\vec\omega.\vec J}$ represents a rotation group element.}
and hence the
contribution of this state to the trace will vanish. For this reason we must
restrict the momentum of the state to be along the directions invariant
under $\vec \omega.\vec J$. The number of such momentum components is
the number $n_T$ introduced in \S\ref{sgrand} (see the discussion in
the paragraph above
\refb{ebp2bmp}). If we assume as in \S\ref{sgrand} that the black hole is
confined in a box of length $L$ along each of these $n_T$ directions then the
trace over the momentum along these directions can be represented, up to
a numerical factor, by
$L^{n_T} \int d^{n_T} P$. 
With this, the statistical grand canonical
partition function $Z$ is related to 
$S_{\rm mc}$ 
via the relation
\ben \label{em1pre}
Z (\beta,  \vec\omega, \vec\mu) 
&\sim& L^{n_T}\, \int d M \, d^{n_T} P \, \sum_{\vec J, \vec Q}
 \, 
e^{-\beta M - \beta (\vec P^2/2M) 
 - \vec\omega . \vec J - \vec \mu
.\vec Q}
e^{S_{\rm mc}( M, \vec J, \vec Q)} \nn
&=& L^{n_T} \int dM\, \left({2\pi M\over \beta}\right)^{n_T/2} \, 
 \sum_{\vec J, \vec Q} \, 
e^{-\beta M - \vec\omega . \vec J - \vec \mu
.\vec Q}
e^{S_{\rm mc}( M, \vec J, \vec Q)}\, .
\een
Clearly the integrand / summand in the right had side of
\refb{em1pre} is sharply peaked around the classical values of
$M,\vec J, \vec Q$ given by the solutions to \refb{elegend}.
Since $\left({2\pi M/ \beta}\right)^{n_T/2}$ is a smooth functions 
of $M,\vec J, \vec Q$, we
can replace it by its classical value and take it out of the integral.
Eqs.\refb{eg4}, \refb{ezscaling} now shows that it scales as
$a^{(D-4)n_T/2}$. Substituting this into \refb{em1pre} and using
\refb{egrfin} we get
\ben\label{em1pre2}
&& \exp\left[\ln Z_{\rm cl}(\beta,  \vec\omega, \vec\mu)+ n_T \ln L+ 
\left( C_{\rm local}  + {1\over 2} n_T (D-4)\right) \, \ln a\right]\nn
&\sim& L^{n_T}\,  a^{(D-4)n_T/2}\, 
\int dM\,  \sum_{\vec J, \vec Q} \, 
e^{-\beta M - \vec\omega . \vec J - \vec \mu
.\vec Q}
e^{S_{\rm mc}( M, \vec J, \vec Q)}\, ,
\een
\i.e.\
\be\label{em1pre3}
\exp\left[\ln Z_{\rm cl}(\beta,  \vec\omega, \vec\mu)+ 
 C_{\rm local}   \, \ln a\right]\nn
\sim 
\int dM\,  \sum_{\vec J, \vec Q} \, 
e^{-\beta M - \vec\omega . \vec J - \vec \mu
.\vec Q}
e^{S_{\rm mc}( M, \vec J, \vec Q)}\, .
\ee
Note that the explicit 
dependence on $n_T$ has cancelled so that the left hand
side (and hence also the right hand side) of this equation has no
discontinuity at special values of $\vec \omega$, {\it e.g.} $\vec \omega=0$,
where there is enhanced rotational symmetry and consequently an increase
in the value of $n_T$.
We can formally invert \refb{em1pre3} to write:
\be \label{em1.1}
e^{S_{\rm mc}( M, \vec J, \vec Q)} \sim
\int \, d\beta\, 
d^{N_C} 
\omega\,  d^{n_V} \mu\, 
\exp\left[\beta M + \vec\omega . \vec J + \vec \mu
.\vec Q + \ln \, Z_{\rm cl} (\beta,  \vec\omega, \vec\mu)
+ C_{\rm local} \ln \, a\right] \, ,
\ee
where $N_C$ is the total number of Cartan generators,
\i.e.\ the rank of the rotation group and $n_V$ is the total number of
Maxwell fields. 
\refb{em1pre3} and \refb{em1.1} are equivalent as long as it is
understood that all the integrals and sums are evaluated 
using saddle point approximation. To leading
order
the location of the saddle point of \refb{em1.1}
is at $(\beta,\vec\omega,\vec\mu)$ satisfying
\be \label{eg2}
M = -{\p \ln Z_{\rm cl}\over \p \beta}, \quad 
J_{b} =-{\p  \ln Z_{\rm cl}\over \p \omega_{b}}, \quad 
Q_\alpha =-{\p  \ln Z_{\rm cl}\over \p \mu_\alpha}\, .
\ee
These relations are the same as those given in \refb{elegend}.
The leading contribution to $S_{\rm mc}$ is
given by $\ln Z_{\rm cl} +\beta M + \vec\omega . \vec J + \vec \mu
.\vec Q$ evaluated at the saddle point, which 
according to \refb{elegend} is the same as
$S_{\rm BH}$. 

In order to study the effect of one loop correction $ C_{\rm local} \ln \, a$
on $S_{\rm BH}$ we can ignore the effect of these
corrections on the saddle
point values of $\beta$, $\vec\omega$ and $\vec\mu$, since
at the leading order
the integrand has been extremized with respect to these variables
at the saddle point, and hence the effect of any change in the
location of the saddle point will affect the result at the second order.
Thus the correction to $S_{\rm mc}$ comes from two sources:
\begin{itemize}
\item $C_{\rm local} \ln a$ evaluated at the saddle point, and
\item the Gaussian integral over $\beta,\vec\omega,\vec\mu$ around the saddle
point.
\end{itemize}
To compute the result of Gaussian integration 
we note that from \refb{ezscaling} we get
\be \label{eg5}
{\p^2 \ln Z_{\rm cl}\over \p \beta^2} \sim a^{D-4}, 
\quad {\p^2 \ln Z_{\rm cl}\over \p 
\omega_{b} \p \omega_{c}} \sim a^{D-2}, \quad
{\p^2 \ln Z_{\rm cl}\over \p 
\mu_\alpha \p \mu_\beta} \sim a^{D-4},
\ee  
and the mixed derivatives scale accordingly. 
If the second
derivative matrix does not have accidental zero
eigenvalues, then
\refb{eg5} allows us to 
express $Z $ as a gaussian peaked around the
saddle point, with the following widths for the different
variables:\footnote{This requires 
the matrix of second 
derivatives to be
negative definite, \i.e.\ the saddle point to be a maximum. However
since Schwarzschild black hole has negative specific heat 
this is not quite
true. Nevertheless we shall proceed by assuming that the
integral can be evaluated using saddle point
approximation by appropriately rotating the integration contours into
the complex plane.}
\ben
\Delta\beta &\sim& a^{-(D-4)/2} \, , \label{ese1} \\
\Delta\omega_{b} &\sim& a^{-(D-2)/2} \, , \label{ese3} \\
\Delta\mu_\alpha &\sim& a^{-(D-4)/2} \, .\label{ese4}
\een
Thus integration over each of these variables will give a contribution
to the right hand side of \refb{em1.1} of this order. 
Combining these results we get
\be \label{eg7}
S_{\rm mc} = S_{\rm BH}+ \ln a \left( C_{\rm local} 
-{1\over 2} (D-4) -{1\over 2}
(D-2) N_C - {1\over 2} (D-4) n_V\right)\, .
\ee

Note that this final formula for $S_{\rm mc}$ is independent of $n_T$.
This can be traced to the cancelation of explicit
$n_T$ dependent factors on the two sides of \refb{em1pre2}.
The absence of $n_T$ dependence in
\refb{eg7} shows that while the grand canonical partition function has
a discontinuity at $\vec \omega=0$ 
where
the black hole has enhanced rotational symmetry,
there is no need for such a discontinuity in $S_{\rm mc}$ at $\vec J=0$.
The discontinuity in the gravitational partition function
$Z$ due to the increased number of zero modes at $\vec \omega=0$
has its counterpart in the statistical partition function due to the explicit
$n_T$ dependent factors on the right hand side of \refb{em1pre} rather
than in any discontinuity in the function $S_{\rm mc}$.
In particular this indicates that the asymptotic expansion
of $S_{\rm mc}(M, \vec J, \vec Q)$ for large charges
can be regarded as 
an analytic function of $\vec J$ even around $\vec J=0$.

\subsection{Entropy in other ensembles} \label{sother}

For spherically symmetric black holes it is often convenient to work in a
mixed ensemble where we keep the charges and mass fixed but sum over
all possible angular momentum states. 
If we denote by $S_{\rm mixed}(M,\vec Q)$ the corresponding entropy then
we have
\be \label{eg7b}
e^{S_{\rm mixed}(M,\vec Q)} 
=  \sum_{\vec J} \, e^{S_{\rm mc}( M, \vec J, \vec Q)
}\, .
\ee
The analog of eqs.\refb{em1pre3}, \refb{em1.1} now take the forms:
\be \label{em1premixed}
\exp\left[\ln Z_{\rm cl}(\beta,  \vec\omega=0, \vec\mu)+ 
 C_{\rm local}   \, \ln a\right]
\sim 
\int dM\,  \sum_{\vec Q} \, 
e^{-\beta M  - \vec \mu
.\vec Q}
e^{S_{\rm mixed}( M, \vec Q)}\, ,
\ee
\be \label{em1.1mixed}
e^{S_{\rm mixed}( M,  \vec Q)} \sim
\int \, d\beta\, 
 d^{n_V} \mu\, 
\exp\left[\beta M +  \vec \mu
.\vec Q + \ln \, Z_{\rm cl} (\beta,  \vec\omega=0, \vec\mu)
+ C_{\rm local} \ln \, a\right] \, .
\ee
Evaluating the integral using saddle point approximation as before we get
\be \label{eg7a}
S_{\rm mixed}(M,\vec Q) = S_{\rm BH}(M, \vec J=0, \vec Q)
+ \ln a \left( C_{\rm local} 
-{1\over 2} (D-4)  - {1\over 2} (D-4) n_V\right)\, .
\ee

We could also consider the other extreme in which we  require
all components of the angular momentum -- not just those
associated with the Cartan generators -- to vanish. This is equivalent to
requiring that we count only singlet states of the rotation group.
The corresponding entropy, denoted as $S_{\rm singlet}(M, \vec Q)$ 
can be calculated as follows. 
Let us denote by $N_R=(D-1)(D-2)/2$ the dimension of the
rotation group, by $\{\chi^1,\cdots \chi^{N_R}\}$ the
parameters labeling a rotation group element (with
$\vec\chi=0$ being the identity element) and by $d^{N_R}\chi$
the Haar measure of the group. Furthermore let 
us denote by $e^{i\vec\theta\cdot \vec T}$ the 
element of the Cartan subgroup
conjugate to $\vec \chi$ with $T^a$'s being the Cartan generator.
Then since $\exp[S_{\rm mc}( M, \vec J, \vec Q)]$ is
the number of states with
eigenvalues $\vec J$ under the Cartan generators, the character of the
representation of the rotation group formed by the black hole
microstates will be given by
\be \label{echar1}
\sum_{\vec J} \exp[S_{\rm mc}( M, \vec J, \vec Q) + i\vec\theta\cdot \vec J]
\, .
\ee
{}From this we can extract the number of singlet states as
\be \label{em1.1newerpre}
\exp[{S_{\rm singlet}( M,  \vec Q)}] =
\int d^{N_R} 
\chi\,  \sum_{\vec J} 
\exp[S_{\rm mc}( M, \vec J, \vec Q) + i\vec\theta\cdot \vec J]\, ,
\ee
where it is understood that $\theta^a$'s are functions of $\vec\chi$.
Now using \refb{em1pre3} we get
\be \label{einter1}
\sum_{\vec J} 
\exp[S_{\rm mc}( M, \vec J, \vec Q) + i\vec\theta\cdot \vec J]
\sim \int \, d\beta\, d^{n_V} \mu\, \exp\left[\beta M + \vec \mu
.\vec Q + \ln \, Z_{\rm cl} (\beta,  \vec\omega=i\vec\theta, \vec\mu)
+ C_{\rm local} \ln \, a\right] \, .
\ee
Using this we can express \refb{em1.1newerpre} as
\be \label{essa1}
\exp[{S_{\rm singlet}( M,  \vec Q)}]\sim
\int \, d\beta\, 
d^{N_R} 
\chi\,  d^{n_V} \mu\, 
\exp\left[\beta M + \vec \mu
.\vec Q + \ln \, Z_{\rm cl} (\beta,  \vec\omega=i\vec\theta, \vec\mu)
+ C_{\rm local} \ln \, a\right] \, .
\ee
We can evaluate this integral using saddle point method. In particular the
saddle point of $\chi$ integration is at the origin. Since near the origin
$\theta^a$'s are degree one homogeneous functions of the coordinates 
$\chi^m$, the integration
over the $N_R$ variables $\{\chi^m\}$ will give a factor of $a^{-N_R(D-2)/2}$ 
according to \refb{ese3}. 
The integration over $\beta$ gives a factor of $a^{-(D-4)/2}$ and integration
over the $\mu_\alpha$'s gives a factor of $a^{-n_V(D-4)/2}$.
Combining these results we get
\be \label{ecas4gen}
S_{\rm singlet}(M, \vec Q) = 
S_{\rm BH}(M, \vec J=0, \vec Q)
+ \ln a \left( C_{\rm local} 
-{1\over 2} (D-4) -{1\over 2}
(D-2) N_R - {1\over 2} (D-4) n_V\right) +\cdots\, .
\ee

\sectiono{Comparison with loop quantum gravity prediction} \label{sloop}

In loop quantum gravity there exist proposals for computing microscopic entropy
of a\break
\noindent Schwarzschild black hole\cite{9505028,9605047,9710007,0005126}.
These results give a formula for the degeneracy as a function of the eigenvalue
of the area operator.
In principle given an exact formula for the degeneracy one can extract its
behaviour for large area by using asymptotic expansion formula. However
there are different versions of this counting formula in loop quantum gravity,
based on $SU(2)$ Chern-Simons theory\cite{9505028,9605047}
and $U(1)$ Chern-Simons theories\cite{9710007,0005126}. 
We shall first review some
of these results and then compare them with the result of the
semi-classical analysis carried out in this paper.

The logarithmic correction to the black hole
entropy based on the SU(2) Chern-Simons theory was first carried out
in \cite{9801080,0002040}, and justified more recently in 
\cite{0905.3168,0907.0846,1006.0634,1103.2723 ,1201.6102}. 
The result for the entropy is given by
\be \label{elogloop}
S^{(lqg)} = S_{\rm BH} - 3\, \ln a\, .
\ee
Before comparing this with our result there
are three important
points to consider:
\begin{enumerate}
\item First of all the analysis 
of \cite{9801080,0002040,0905.3168,0907.0846,1006.0634,1103.2723 ,1201.6102} 
counted all states carrying
a fixed number $p_0$ of `punctures'. This number is an integer and is
related to the area via the relation
\be \label{epuncarea}
p_0 \propto A_H \, .
\ee
Since $A_H\sim M^2$ we have $\delta p_0 \sim M \, \delta M$. Now since
$p_0$ is integrally quantized the counting 
of \cite{9801080,0002040,0905.3168,0907.0846,1006.0634,1103.2723 ,1201.6102} gives the
number of microstates per unit $p_0$ interval. This corresponds to
$\delta M \sim 1/M$. Thus in order to get the number of 
states per unit mass interval, we need to multiply the number of
states counted in \cite{9801080,0002040,0905.3168,0907.0846,
1006.0634,1103.2723 ,1201.6102} 
by $M\sim a$. This gives an additional
contribution of $\ln a$ to the entropy. 

\item
Since the presence of logarithmic correction to the effective action could
change the equations of motion and hence
the relation between mass and area, this could
give additional logarithmic corrections to the entropy when we express
the leading result in loop quantum gravity -- given by a term
proportional to the area -- in terms of the mass of the
black hole. This can certainly
happen for a general charged black hole, but does not happen
for a Schwarzschild black hole for the following reason.
For Schwarzschild
black hole the $K_0(x)$ term differs from a multiple of the Gauss-Bonnet term
by $R_{\mu\nu}R^{\mu\nu}$ and $R^2$
terms. Both these terms vanish on-shell, and furthermore their contribution to the
equation of motion, being proportional to the first variation of these
terms, also vanish on-shell. Finally 
the Gauss-Bonnet term being total
derivative also does not contribute to the equations of motion.
{}From this it follows
that the Schwarzschild black hole solution does not receive any
logarithmic correction. There can be corrections obtained by varying the
$\ln a$ term multiplying $K_0(x)$ since $a$ can be expressed
in terms of the metric, but since $\delta \ln a=\delta a/a$,
this will not be a logarithmic correction to the
equations of motion. Thus we conclude that for the Schwarzschild black
hole in $D=4$, there is no additional logarithmic correction to the result
\refb{elogloop} due to a
modification of the relation between mass and area of the horizon.

\item \cite{9801080,0002040,0905.3168,0907.0846,1006.0634,1103.2723 ,1201.6102} 
counts all spherically symmetric states,
\i.e.\ all states of zero angular momentum.
Thus it gives the result for $S_{\rm singlet}$ described in \S\ref{sother}.
\end{enumerate}
The upshot of this discussion is that the loop quantum gravity prediction
for $S_{\rm singlet}$ is given by adding to \refb{elogloop} a term $\ln a$.
This gives
\be \label{elogloop1}
S^{(lqg)}_{\rm singlet} = S_{\rm BH} -2\, \ln a\, .
\ee

An alternative computation of the
corrections to the Schwarzschild black hole entropy in loop quantum
gravity, based on the U(1) Chern-Simons theory, 
has been given in \cite{0006211,0407052,0411035,0605125}. 
This also counts states with a fixed number of punctures and arrives
at the result  $-\ln a$ for logarithmic correction to the
entropy. 
The difference between this and the computation based on SU(2)
Chern-Simons theory can be traced to the different projections used
in the two computations under the global part of the SU(2) Chern-Simons
gauge group\cite{0907.0846,1201.6102}. While the analysis of 
\cite{9801080,0002040,0905.3168,0907.0846,1006.0634,
1103.2723 ,1201.6102} imposes the constraint that the states are singlets of the
global part of the SU(2) gauge group, 
\cite{0006211,0407052,0411035,0605125}  only requires invariance under
a U(1) subgroup of this SU(2) group. If we identify the global part of the
SU(2) group as the rotational isometry of the black hole\cite{1201.6102}, 
then
the first one counts states with $\sum_i J_i J_i=0$ while the second one
counts states with $J_3=0$ but arbitrary $\sum_i J_i J_i$. 
In other words, the second result, after being converted to the number of
states per unit mass range by adding $\ln a$, gives the entropy $S_{\bf mc}$
in loop quantum gravity:
\be \label{elogloop12}
S^{(lqg)}_{\rm mc} = S_{\rm BH}\, .
\ee
Note that there are no logarithmic corrections to $S_{\rm mc}$ in
loop quantum gravity.
This agrees with \refb{elogloop1}
since fixing $J_3$ to 0
but letting $\sum J_i J_i$ to be
arbitrary produces an additive factor of $2\ln a$ in the entropy compared
to the case where we fix both $J_3$ and $\sum J_i J_i$ to 0
(compare \refb{eg7}
and \refb{ecas4gen} for $D=4$, $N_C=1$, $N_R=3$).

Let us compare \refb{elogloop1} with the prediction from Euclidean gravity
analysis.
Since we are considering a theory of pure gravity, we have,
from \refb{ecas4gen}, \refb{eclocfin} with $n_S=n_V=n_F=n_{3/2}=0$,
\be \label{elogsch}
S_{\rm singlet} =S_{\rm BH} + (C_{\rm local}-3) \ln a, \quad C_{\rm local} = {212\over 45}
\, .
\ee
This is different from \refb{elogloop1}, showing that the loop quantum
gravity result for logarithmic correction to the entropy does not agree
with the prediction of the Euclidean gravity analysis.

\sectiono{BTZ black holes} \label{sbtz}

The logarithmic corrections to the
entropy of BTZ black holes have been computed both from the
microscopic\cite{0005017} and macroscopic\cite{0005003,0104010,0111001}
perspective. In this section we shall see
how the formalism described in this paper can be used to compute
these corrections. 
This section does not contain any new results, but simply
translates the existing analysis into the framework used in this
paper for computing logarithmic corrections to the entropy.
We shall first review the microscopic computation
and then describe the macroscopic computation.

\subsection{Microscopic computation} \label{smicrobtz}

We shall consider a (1+1) dimensional conformal field theory with
central charges $(c, \bar c)$. If
$d_0(n,\bar n)$ denotes the degeneracy of
states carrying $(L_0,\bar L_0)$ eigenvalues $(n,\bar n)$, then
we define the partition function as
\be \label{ebtz2pre1}
Z(\tau, \bar\tau) =
Tr\left[ e^{2\pi i \tau L_0 - 2\pi i \bar\tau \bar L_0}\right]
= \sum_{n, \bar n} \,
d_0(n,\bar n)
 \, e^{2\pi i n \tau - 2\pi i \bar n \bar\tau } \, .
\ee
In this sum $n,\bar n$ are typically discrete but not necessarily integers,
although $n-\bar n$ takes integer values. For small $\tau,\bar\tau$ the
contribution to the integral comes from large $n$, $\bar n$. In this case
we can approximate the sum by an
integral of the form
\be \label{ebtz2pre}
Z(\tau, \bar\tau) \simeq \int dn \int d\bar n  \,
d(n,\bar n)
 \, e^{2\pi i n \tau - 2\pi i \bar n \bar\tau } \, ,
\ee
where $d(n,\bar n)$ is some smooth function representing the average
number of states per unit interval in $n$ and $\bar n$. Note that
$d(n,\bar n)$ could differ from $d_0(n,\bar n)$ by a large factor if
the spacing between $L_0+\bar L_0$ eigenvalues is small.
Now
modular invariance of the theory implies that
\be \label{ebtz3}
Z(\tau, \bar\tau)
= Z\left(-{1\over \tau},  -{1\over\bar \tau}\right)\nn
= Tr\left[ e^{-2\pi i {1\over\tau} L_0  + 2\pi i {1\over\bar\tau} \bar L_0}
\right]\, . 
\ee
For small $\tau$, $\bar \tau$ the contribution to the right hand side is
dominated by the vacuum state with $L_0=-{c\over 24}$,  
$\bar L_0=-{\bar c\over 24}$ and we have
\be \label{ebtz3.5}
 Z(\tau, \bar\tau) 
 \simeq \exp\left[ \pi i {c\over 12\tau} -  \pi i {\bar c\over 12\bar \tau}
\right] \, .
 \ee
This gives, for large $n,\bar n$
\be \label{ebtz4}
d(n, \bar n) \simeq \int \, d^2\tau 
e^{-2\pi i n\tau + 2\pi i \bar n\bar \tau} Z(\tau, \bar\tau) 
\simeq \int d^2\tau 
e^{-2\pi i n\tau + 2\pi i \bar n\bar \tau+ \pi i {c\over 12\tau} 
- \pi i {\bar c\over 12\bar \tau}} \, , \ee
where it is understood that on the right hand side we pick the
contribution to the integral from the saddle point close to the origin.
Now the
saddle point, obtained by extremizing the exponent on the right hand side
of \refb{ebtz4},  is at 
\be \label{ebtztau0}
\tau_0= i \sqrt{c\over 24 n},
\quad
\bar\tau_0= -i \sqrt{\bar c\over 24\bar n},
\ee
 and the
result of the integration is given by
\ben \label{ebtz6}
d(n, \bar n) &\simeq& \exp\left[ 2\pi \sqrt{{cn\over 6}
}
+ 2\pi \sqrt{{\bar c\bar n\over 6}}
\right] (-12\tau_0^3/i c)^{1/2} (12\bar\tau_0^3/i \bar c)^{1/2}\nn
&\simeq& C_0 \exp\left[ 2\pi \sqrt{{cn\over 6}
} 
+ 2\pi \sqrt{{\bar c\bar n\over 6}
}  -
{ 3\over 4} \ln n
-
{3\over 4} \ln \bar n\right]
\een
for some constant $C_0$.

\subsection{Macroscopic computation} \label{smacrobtz}

The three dimensional theory in the bulk that is dual to the
$CFT_2$ described in \S\ref{smicrobtz} is 
a theory of gravity in $AdS_3$ space-time
with Einstein-Hilbert term, cosmological constant term
and gravitational Chern-Simons term, with the constants $c$ and
$\bar c$ given by specific  combinations of the parameters of the
bulk theory:
\be \label{egcs2}
c+\bar c= {3 l\over G}, \quad c-\bar c = 48\pi K\, ,
\ee
where
$l$ is the radius of curvature of the dual $AdS_3$ space,
$G$ is the Newton's constant and $K$ is proportional to the
coefficient of the gravitational Chern-Simons term.
$\ln d(n, \bar n)$ has to be compared with the microcanonical
entropy
of a BTZ black hole of mass $M$ and angular momentum $J$
with the identification
\be \label{eidenbtz}
Ml=n+\bar n, 
\qquad J=n-\bar n \, .
\ee

Now
the Bekenstein-Hawking entropy of a BTZ black hole\cite{9204099}
carrying mass $M$ and angular momentum $J$, is given by
(see {\it e.g.} \cite{0506176})
\be \label{esbhbtz}
S_{\rm BH}=\pi \sqrt{{c\over 3}
\left(Ml+J\right)} 
+ \pi \sqrt{{\bar c\over 3}
\left(Ml-J\right)} = 2\pi \sqrt{c\, n\over 6} + 2\pi\sqrt{\bar c \, \bar n\over 6}
\, .
\ee
This is
in perfect agreement with the leading terms in \refb{ebtz6}. 
Our goal will be to compute the logarithmic correction
to the entropy from the macroscopic side. For this we need to first
determine the scaling laws of various quantities with the size of the
black hole.
If $a$ denotes the size of the black hole horizon then we have
$S_{\rm BH} \sim a/G_N$. Comparing this with \refb{esbhbtz} we see that
here
\be \label{escalebtz}
M\sim a^2, \quad J\sim a^2, \quad S_{\rm BH}\sim a\, .
\ee
Since we have a cosmological constant, it follows from the arguments
at the end of \S\ref{sgrand} that the non-zero 
modes do not produce any logarithmic
correction to the partition function. Furthermore none of the spatial
translation
generators commute with the generator of rotation in the two dimensional
plane and hence there are no translational zero modes of the Euclidean
black hole solution. As a result there are no logarithmic corrections to
$\ln Z$, and the only logarithmic corrections to the entropy come from
the conversion of the grand canonical partition function to
microcanonical entropy via the relation
\be \label{econvbtz}
Z(\beta,\omega) = \int dM \, \sum_J \, e^{S_{\rm mc}(M,J)-\beta M
-\omega J}\, ,
\ee
or equivalently
\be \label{ebtzrev}
e^{S_{\rm mc}(M,J)} = \int d\beta d\omega \, Z(\beta,\omega) e^{\beta M
+\omega J}\, .
\ee
It follows from \refb{escalebtz} that 
we have $\beta\sim a^{-1}$, $\omega\sim a^{-1}$, $\ln Z_{\rm cl}
\sim a$ and hence $\p^2 \ln Z_{\rm cl}/\p \beta^2 \sim
\p^2 \ln Z_{\rm cl}/\p \omega^2 \sim \p^2 \ln Z_{\rm cl}/\p \beta\p \omega 
\sim a^{3}$. Thus
integrations over $\beta$ and $\omega$ together
produces a factor of $a^{-3}$ and gives
\be \label{ezsmi}
e^{S_{\rm mc}} \sim \left[Z \, e^{\beta M+\omega J}\right]_{\rm saddle} \, a^{-3}\, .
\ee
Absence of logarithmic correction to $Z$ now gives the logarithmic
correction to $S_{\rm mc}$ to be 
\be \label{egives}
S_{\rm mc} = S_{\rm BH} - 3\ln a\, .
\ee

\subsection{Comparison of the microscopic and the macroscopic results}
\label{scompbtz}

To compare the microscopic and the macroscopic results we note that
\refb{eidenbtz} and \refb{escalebtz} together gives
\be \label{ennscaling}
n\sim a^2, \quad \bar n\sim a^2\, .
\ee
Thus we can express the microscopic result \refb{ebtz6} as
\be \label{esslog}
\ln d(n,\bar n) = S_{\rm BH} 
-3\, \ln a\, .
\ee
This is in perfect
agreement with the macroscopic result \refb{egives}.

Note that both the macroscopic and the microscopic results
\refb{egives} and \refb{esslog} 
differ from the
result $-{3\over 2}\ln a$ given in \cite{0005017} by a factor of 2. This
can be traced to the fact that in \cite{0005017} the entropy was calculated
in a mixed ensemble in which the mass $M=n+\bar n$ was fixed but the
angular momentum $J=n-\bar n$ was summed over. For this we need to
set $\tau+\bar\tau=0$ and only integrate over $\tau-\bar\tau$ in
eq.\refb{ebtz4}. Evaluating the integral by saddle point method we shall
get the result $-{3\over 2}\ln a$.
Similarly on the macroscopic side the computation of the entropy in
the mixed ensemble will involve a relation like \refb{ebtzrev} with 
$S_{\rm mc}(M,J)$ replaced by $S_{\rm mixed}(M)$ on the left hand side,
and 
$\int d\beta e^{\beta M} Z(\beta,\omega=0)$ on the right hand side.
The integration over $\beta$ will now produce
a factor of $a^{-3/2}$, giving us a logarithmic correction of $-(3/2)\ln a$
to $S_{\rm mixed}$.

Finally we note that the agreement between the logarithmic corrections in the
microscopic and macroscopic analysis could also have been inferred by
comparing the gravity partition function $Z$ with the conformal field
theory partition function $Z(\tau,\bar\tau)$ given in \refb{ebtz3.5}.
The absence of logarithmic corrections to both the gravity partition
function and the conformal field theory partition function is enough to
ensure that the corresponding microcanonical entropies should also
agree.

\bigskip

{\bf Acknowledgement:} I would like to thank  A.~Ghosh, R.~Kaul, Cynthia Keeler,
P.~Majumdar, K.~Meissner and P.~Mitra
for useful communications. This work was
supported in part by the J. C. Bose fellowship of 
the Department of Science and Technology, India and the 
project 11-R\&D-HRI-5.02-0304.

\appendix

\sectiono{Normalizability of the zero modes}  \label{sa}

A black hole solution breaks translation invariance, and, if it carries
angular momentum, also breaks the rotational invariance to its Cartan
subalgebra. Thus naively one would expect the solution to carry zero mode
deformations associated with broken translational and rotational invariance.
We shall however show that for the euclidean black hole
only the zero modes associated with broken
translation symmetry, which commute with the rotation
generator $\vec\omega\cdot \vec T$, satisfy the required boundary conditions.
All other zero modes are projected out.

Let us first analyze the zero modes of the Lorentzian black hole
associated with broken translation
invariance. These are generated by diffeomorphisms which
approach constant translation at infinity and vanish as we
approach the horizon. Since for testing normalizability we only need
the behaviour of the deformation at infinity, we can work with
constant translations.
We shall begin by examining Schwarzschild solution in
$D$ dimensions. The metric takes the form\cite{myersperry}
\be \label{esa1}
ds^2 = -(1 - C r^{-D+3}) dt^2 + (1 - C r^{-D+3})^{-1} dr^2 + r^2 d\Omega_{D-2}^2
\, ,
\ee
where $d\Omega_d$ is the line element on a unit $d$-sphere.
If we introduce the Cartesian coordinates $(x^0=t,x^1,\cdots x^{D-1})$
in the usual manner and deform the solution by a diffeomorphism
that approaches, for $r\to\infty$,
$\delta x^i=a^i$ for some fixed $(D-1)$ dimensional vector $\vec a$, then
$\eta_{\mu\nu} dx^\mu
dx^\nu$ -- 
the $C$ independent part of the metric -- 
remains invariant under this transformation. The change in
the $C$ dependent part can be computed using the fact that under this
transformation $\delta r= \vec a\cdot \vec x / r$. Thus we have
\ben \label{esa2}
\delta (ds^2) &=& C (3-D) \vec a\cdot \vec x \, r^{1-D} dt^2 
+ (1 - C r^{-D+3})^{-2} C (3-D) \vec a\cdot \vec x \, r^{1-D} dr^2\nn
&&
+ 2\left((1 - C r^{-D+3})^{-1}  -1 \right) \left(r^{-1} \vec a\cdot d\vec x 
- r^{-2} \, \vec a\cdot \vec x\, dr
\right) dr\, ,
\een
to first order in the deformation parameter $\vec a$.
This shows that in the asymptotically Cartesian coordinate system
$\delta g_{\mu\nu}$ is of order $r^{2-D}$. 
These are clearly normalizable deformations of the solution for $D\ge 4$.
For rotating black holes the metric is more complicated but asymptotically
the metric approaches the Minkowski metric at the same rate. Thus the
translation zero modes are normalizable for these solutions as well.

Let us now analyze the fate of these zero modes 
when we consider Euclidean (rotating)
black holes obtained by the replacement $t\to -i\tau$ followed
by periodic
identification of the $\tau$ coordinate.
Let us for definiteness focus on the  
Kerr or Kerr-Newmann metric in four dimensions, but the analysis 
generalizes to higher dimensions. The metric near the horizon has a 
$(d\phi+i\omega  d\tau/\beta)^2$ factor which  
becomes singular at the horizon if we make the usual
identification $\tau\equiv \tau+\beta$ and $\phi\equiv \phi+2\pi$.
The remedy is to define $\wt\phi=\phi +i\omega\tau/\beta$ and make the
identification $\wt\phi\equiv\wt\phi+2\pi$ and $\tau\equiv\tau+\beta$. In
terms of the original $(\phi,\tau)$ coordinate system this corresponds to
an identification $(\phi,\tau)\equiv (\phi-i\omega, \tau+\beta)\equiv
(\phi+2\pi,\tau)$.
Thus the question we need to address is whether the translation zero modes
satisfy this periodicity restriction. From the description of the translation
zero modes given above it is clear that under $\tau\to\tau+\beta$
the parameters $\vec a$ remain unchanged, whereas under 
$\phi\to \phi-i\omega$, $a_z\to a_z$, $(a_x+ia_y)\to
e^{\omega\beta} (a_x+ia_y)$. Thus the zero mode generated by the
parameter $a_z$ satisfies the required boundary condition while the
zero modes generated by $a_x$ and $a_y$ fail to satisfy the
requirement of periodicity under 
$(\phi,\tau)\to (\phi-i\omega, \tau+\beta)$. 
For more general black holes in higher dimensions one
can generalize this analysis to argue that only those translations which
commute with the rotation generator
$\vec\omega.\vec T$ generate zero modes. For 
$\vec\omega=0$, \i.e.\ for Schwarzschild black holes, all the translation
generators produce zero modes of the Euclidean black hole solution.

We now turn to the rotational zero modes. In this case asyptotically
the black hole metric deviates from the rotationally  invariant
metric by a term
of order $dt dx^i / r^{D-2}$. On the other hand under a rotation
$\delta x^i\sim r$. Thus
the asymptotic form of $\delta g_{ti}$ associated with a rotational zero mode
is of order $1/r^{D-2}$ and again this describes a normalizable deformation
for $D\ge 4$.
As a result for Lorentzian rotating black holes there are zero modes
for every broken rotation generator. However things are again different in the
Euclidean theory. Following the logic of the previous two subsections we
can show that all the rotational zero modes transform non-trivially under
$\vec\omega.\vec T$ and hence fail to satisfy the periodicity requirement
imposed by the Euclidean solution. Thus for the euclidean
black hole there are
no rotational zero modes.

\sectiono{Subtraction of the thermal gas contribution to the partition function}
\label{sb}

In this appendix we shall identify the thermal gas contribution to the
black hole partition function and argue that the subtraction scheme
used in the text, in which we divide the partition function $Z$
of the original
black hole of size $a$ and confined in a box of size $L$
by the partition function $Z_0$ of a black hole of fixed size $a_0$
and confined in a box of size $La_0/a$, correctly removes not only
the
contribution to  $Z$ from the thermal gas but also the 
spurious boundary
contributions.

In order to gain some insight into the problem, it will be useful to
recall how the euclidean partition function in the background of
a flat space-time in which the euclidean time coordinate has
period $\beta$ sees the thermal contribution to the
partition function. 
In this case the  eigenvalues of the kinetic operator
have the form:
\be \label{eb1}
{4\pi^2 n^2 \over \beta^2} + \vec k^2\, ,
\ee
with density of states
\be \label{eb2}
d\mu = {V\over (2\pi)^{D-1}} d^{D-1} k + \cdots\, ,
\ee
where $V\sim L^{D-1}$ is the total volume of the box inside which we
restrict the spatial coordinates and
$\cdots$ denote subleading corrections involving lower powers of
$L$ which arise from boundary effects. Thus the net contribution
to the partition function -- which we shall denote by $Z_{\rm free}$ --
is given by
\be \label{eb3}
\ln Z_{\rm free}(\beta, L, \eps)=
{1\over 2} \, \int_\eps^\infty {ds\over s} \sum_{n=-\infty}^\infty
\int \, d\mu \, e^{-4\pi^2 n^2 s/\beta^2 - s \vec k^2}\, .
\ee
We now perform the sum over $n$ by Poisson resummation 
formula
\be \label{eb4}
\sum_{n=-\infty}^\infty f(n) = \sum_{m=-\infty}^\infty \wt f(m), 
\qquad \wt f(m) = \int dx \, e^{-2\pi i m x} f(x)\, ,
\ee
to
express \refb{eb3} as
\be \label{eb4new}
\ln Z_{\rm free}=
{1\over 2} \, \int_\eps^\infty {ds\over s} \sqrt{\beta^2\over 4\pi s}\,
\sum_{m=-\infty}^\infty
\int \, d\mu \, e^{- m^2\beta^2 / 4 s - s \vec k^2}\, .
\ee

First consider the contribution from the
$V \, d^{D-1}k / (2\pi)^{D-1}$ term in $d\mu$. 
The $m=0$ term is ultraviolet divergent, but it is independent of
$\beta$ except for an overall multiplicative factor of $\beta$.
This has the interpretation of one loop contribution to the cosmological
constant and will be 
cancelled by a local counterterm if the final theory does not have
a cosmological constant. 
The $m\ne 0$ terms have their integrand suppressed
exponentially for $s<<a^2$ and hence are free from ultraviolet 
divergences. For these terms we can set $\eps=0$ and it is easy
to see that the result will be proportional to $V \beta^{-D+1}$.
This is precisely the thermal contribution
to the partition function. 

Next consider the effect of the boundary
corrections to the density of states, represented by $\cdots$ in
\refb{eb2}. It will be important to understand the nature
of these contributions since  the
analog of such contributions for the black hole problem must also be
removed from the partition function if we are to compute the
partition function associated with black hole microstates. 
Like the bulk contribution, the
$m=0$ term may be ultraviolet divergent, \i.e.\ divergent in the $\eps\to 0$
limit. Since the $\eps$ dependence of the contribution can be infered from
the small $s$ expansion of the heat kernel, which naturally splits into
a bulk and the boundary contribution, we can remove the ultraviolet
divergent part of the boundary contribution by simply removing from the
heat kernel the boundary contribution to the 
non-positive powers of $s$ in the
small $s$ expansion. As explained in
the second point in the discussion below \refb{eclocal}, this is the prescription
used in \S\ref{sgrand} for dealing with the boundary contribution to the
heat kernel. 
The rest of the contribution, being $\eps$ independent, now depends only
on $L$ and $\beta$. By dimensional analysis  this must depend on
$L/\beta$. Thus if we subtract from this
$\ln Z_{\rm free}(\beta_0, L_0, \eps)$ with $L_0=L\beta_0/\beta$
then
the boundary contributions will cancel altogether irrespective of their
form. (In this case the bulk contributions, representing the
thermal partition function, also cancel completely, 
but this will not be the case
for the black hole.)

Let us now consider the effect of the presence of the black hole inside
the box. In this case 
we do not have a factorization between the 
contribution from the modes along the time circle and the modes along the
spatial directions.
As a result
the 
eigenvalues of the kinetic operator have more complicated form than
\refb{eb1}, and the
density of states depends on the quantum
number $n$ labelling momentum along the Euclidean time circle.
Also since
for large distance
away from the black hole the spatial part of the kinetic operator
resembles the Hamiltonian of a charged particle in a Coulomb field,
the eigenfunctions are not plane waves but have additional phases
which depend on the radial coordinate. 
Nevertheless our general arguments hold.
In
particular 
in the large $L$ limit the dominant contribution to $\ln Z$,
proportional to $L^{D-1}$, is expected to be identical to that for
a thermal gas without a black hole, and is cancelled against a
similar factor in the expression for $\ln Z_0$. For the terms
containing subleading powers of $L$,
once we remove from the short distance expansion of the
heat kernel the boundary terms containing non-positive powers of
$s$, there is no ultraviolet divergent contribution associated with the
boundary and the $L$-dependent contribution depends
only on the ratio $L/a$. Since $L_0/a_0=L/a$, 
subtracting $\ln Z_0$ from $\ln Z$ also removes all possible
subleading
boundary contributions which depend on $L/a$.  This is reflected in the
fact that the contribution \refb{e4} to $\ln (Z/Z_0)$ can be evaluated without
any knowledge of the box in which we confine the black hole, as long
as we use the prescription of dropping the
non-positive powers of $s$ in the 
boundary contribution to the heat kernel.

\end{document}